\documentclass[twocolumn]{aastex61}
\pdfoutput=0

\usepackage{subfigure}
\usepackage{url}
\usepackage{comment}
\usepackage{color}
\usepackage{enumitem}
\usepackage{amsmath}
\usepackage[section]{placeins}
\usepackage{graphicx,booktabs}

\newcommand{\cms}[1]{{\color{purple}{[\;\bf #1\;]}}}
\newcommand{\new}[1]{{\color{black}{#1}}}

\setlist[enumerate]{noitemsep}
\setlist[enumerate,1]{leftmargin=*}
\setlist[itemize]{noitemsep}
\setlist[itemize,1]{leftmargin=*}
\setlist[description]{noitemsep}
\setlist[description,1]{leftmargin=*}
\setenumerate[0]{label=(\arabic*)}
\shorttitle{M15}\shortauthors{Cabrera Garcia et al.}

\begin{document}
\title{Abundances of Neutron-Capture Elements in 62 Stars in the Globular Cluster Messier 15}

\author{Jonathan Cabrera Garcia}
\affil{Department of Physics and Astronomy and JINA Center for the Evolution of the Elements, University of Notre Dame, Notre Dame, IN 46556, USA}
\affil{Department of Physics \& Astronomy, San Francisco State University, San Francisco CA 94132, USA}

\author{Charli M. Sakari}
\affil{Department of Physics \& Astronomy, San Francisco State University, San Francisco CA 94132, USA}

\author{Ian U.\ Roederer}
\affil{%
Department of Physics, North Carolina State University,
Raleigh, NC 27695, USA}
\affil{%
Department of Astronomy, University of Michigan,
Ann Arbor, MI 48109, USA}
\affil{%
Joint Institute for Nuclear Astrophysics -- Center for the
Evolution of the Elements (JINA-CEE), USA}

\author{Donavon W. Evans}
\affil{Department of Physics \& Astronomy, San Francisco State University, San Francisco CA 94132, USA}

\author{Pedro Silva}
\affil{Department of Physics \& Astronomy, San Francisco State University, San Francisco CA 94132, USA}

\author{Mario Mateo}
\affil{%
Department of Astronomy, University of Michigan,
Ann Arbor, MI 48109, USA}

\author[0000-0002-6270-8851]{Ying-Yi Song}
\affiliation{David A. Dunlap Department of Astronomy \& Astrophysics, University of Toronto, 50 St. George Street, Toronto, ON M5S 3H4, Canada}
\affiliation{Dunlap Institute for Astronomy \& Astrophysics, University of Toronto, 50 St. George Street, Toronto, ON M5S 3H4, Canada}


\author{Anthony Kremin}
\affil{%
Physics Division, Lawrence Berkeley National Laboratory,
Berkeley, CA 94720, USA}

\author[0000-0002-4272-263X]{John I.\ Bailey, III}
\affil{%
Department of Physics, University of California,
Santa Barbara, CA 93106, USA}




\author{Matthew G.\ Walker}
\affil{%
Department of Physics, Carnegie Mellon University,
Pittsburgh, PA 15213, USA}

\correspondingauthor{Charli M. Sakari}\email{sakaricm@sfsu.edu}

\begin{abstract}
M15 is a globular cluster with a known spread in neutron-capture elements.  This paper presents abundances of neutron-capture elements for 62 stars in M15.  Spectra were obtained with the Michigan/Magellan Fiber System (M2FS) spectrograph, covering a wavelength range from $\sim$4430-4630 \AA.  Spectral lines from \ion{Fe}{1}, \ion{Fe}{2}, \ion{Sr}{1}, \ion{Zr}{2}, \ion{Ba}{2}, \ion{La}{2}, \ion{Ce}{2}, \ion{Nd}{2}, \ion{Sm}{2}, \ion{Eu}{2}, and \ion{Dy}{2}, were measured, enabling classifications and neutron-capture abundance patterns for the stars.  Of the 62 targets, 44 are found to be highly Eu-enhanced $r$-II stars, another 17 are moderately Eu-enhanced $r$-I stars, and one star is found to have an $s$-process signature.  The neutron-capture patterns indicate that the majority of the stars are consistent with enrichment by the $r$-process.  The 62 target stars are found to show significant star-to-star spreads in Sr, Zr, Ba, La, Ce, Nd, Sm, Eu, and Dy, but no significant spread in Fe.  The neutron-capture abundances are further found to have slight correlations with sodium abundances from the literature, unlike what has been previously found; follow-up studies are needed to verify this result.  The findings in this paper suggest that the Eu-enhanced stars in M15 were enhanced by the same process, that the nucleosynthetic source of this Eu pollution was the $r$-process, and that the $r$-process source occurred as the first generation of cluster stars was forming.
\end{abstract}

\section{Introduction}\label{sec:Intro}
Metal-poor stars have a metallicity $[\text{Fe/H}] < -1$ \citep{BeersChristlieb2005} and are generally much older than the Sun.   The more metal-poor the star, the less it has been enriched by the generations of stars that came before.  A handful of metal-poor stars are known to be so metal-poor that their atmospheres are promising laboratories that resemble the environment of the early universe, before many metals were formed but after the first-generation of  massive stars had evolved and died (e.g., \citealt{Keller2014,Frebel2019}). The chemical abundances of the most metal-poor stars are therefore particularly useful for understanding the conditions of the early Universe.  

The periodic table is an organized grouping of the known elements existing in our universe. Early massive stars fused heavy elements in their cores, eventually reaching iron.  Exploding stars and their remnants enrich primordial gas clouds, which become the birth places for more stars to form.  Subsequent generations of stars then become increasingly enriched in iron.  For elements heavier than the iron-group elements, exploding massive stars, dying low mass stars, neutron star mergers (NSMs), \new{and black hole-neutron star mergers} are thought to be potential astrophysical sites to manufacture these elements via the neutron-capture process (see, e.g., the review by \citealt{Frebel2018}). This process involves seed nuclei (e.g. Fe) capturing free neutrons within a neutron-rich environment, followed by beta decays.  The neutron-capture processes are often divided into two categories: the slow neutron capture process ($s$-process) and the rapid neutron capture process ($r$-process).  For heavy elements to be produced via the $r$-process, seed nuclei must enter an environment with high neutron density of $n_{n} > 10^{22} \, \text{cm}^{-3}$ so that the nuclei can collect many neutrons forming unstable isotopes inevitably decaying into stable isotopes such as thorium ($Z=90$) and uranium ($Z=92$; \citealt{Frebel2018}).  

The astrophysical site(s) of $r$-process nucleosynthesis remain somewhat mysterious.  Based on follow-up observations of the gravitational wave event from a binary neutron star inspiral \citep{Abbott2017}, NSMs are one site known to produce the heaviest elements via the $r$-process \citep{Chornock2017,Drout2017,Shappee2017}. NSMs are a promising site for explaining many observations of metal-poor stars \citep{Holmbeck2021}, but it is still not clear whether NSMs are the main astrophysical sites for the $r$-process, particularly at early times (\citealt{Tsujimoto2015,Tsujimoto2017}; \new{\citealt{Kobayashi2023})}.  Continued observations of metal-poor stars, including $r$-process enhanced metal-poor stars, provide an excellent opportunity to obtain pristine measurements of heavy elements formed early on by the $r$-process (e.g., \citealt{Hansen2018,Sakari2018,Ezzeddine2020,Holmbeck2020}).

Metal-poor stars can also be found in globular clusters (GCs).   GCs are collections of tens of thousands of stars that are gravitationally bound. GCs were once thought to have formed from giant molecular clouds with a homogeneous chemical composition; as a consequence, they were believed to exhibit star-to-star consistency in their chemical abundances.  Recent studies have shown this is not generally true (e.g., \citealt{Carretta2009,Gratton2012}). For instance, it is found that there are spreads of heavy elements, including neutron-capture elements and even iron, in GCs \citep{Roederer2011, Johnson2010, Johnson2017}. Most GCs have a moderate enhancement in the $r$-process \citep{Gratton2004}, \new{though one, NGC~5986, has been found with significant Eu enhancement amongst all stars \citep{Johnson2017_Euhigh}}.  Star-to-star spreads in the $r$-process could suggest progenitor events like NSMs being present during the formation of GCs. The formation of GCs is still not well understood, thus probing the atmospheres of metal-poor stars in GCs and investigating the chemical abundances of neutron capture elements such as barium and europium could shed light behind galaxy formation, galaxy structure, and the rates of NSMs during the epoch of GC formation. 

M15, also known as NGC 7078, is one of the only GCs known to have a significant star-to-star spread in $r$-process elements (e.g., \citealt{Sneden1997,Sneden2000a,Sneden2000b,Otsuki2006,Sobeck2011,Worley2013}), \new{hosting metal-poor stars with strong Eu enhancement, similar to the highly enhanced $r$-II stars in the Milky Way, as well as stars that show only a moderate Eu-enhancement.  Unlike Milky Way halo stars and $r$-process-enhanced stars in other environments, like the dwarf galaxy Reticulum~II \citep{Ji2016,Roederer2016b}, M15 places important timing constraints on the $r$-process event.  \citet{Kirby2020} show that the nucleosythetic source of the $r$-process elements had to have happened before the cluster stars had finished forming.  M15's old age (e.g., \citealt{Monelli2015,Vandenberg2016}) furthermore requires the $r$-process event to occur early in the Universe.  These criteria suggest that an $r$-process nucleosynthetic event occurred as the GC was forming.} This makes M15 an excellent source in the Galaxy to observe and further understand $r$-process nucleosynthesis. 

\new{The $r$-process spreads within M15 have been explained by prolonged star formation within the GC that leads to an $r$-process event within the GC \citep{BekkiTsujimoto2017,Zevin2019} or a serendipitous $r$-process event near the cluster \citep{Tarumi2021}.  Such scenarios can be constrained based on the nature of the $r$-process spreads and their relationships with the light element spreads within the cluster (e.g., \citealt{Carretta2009}).  Recently, \citet{Kirby2023} found evidence for $r$-process spreads in another metal-poor GC, M92, but only in the stars with low sodium (which are sometimes referred to as ``first-generation'' GC stars).  Although the site of light element variations within GCs is still debated (see \citealt{BastianLardo2018}), they are ubiquitous within classical Milky Way GCs (e.g., \citealt{Carretta2009}).  \citet{Kirby2023} argue that the relationship between the $r$-process and light elements indicates that the $r$-process event happened as M92 was forming and could therefore not be caused by a NSM within the cluster.  Although previous papers have not found a similar relationship between the neutron-capture and light elements in M15 \citep{Roederer2011}, these findings have only come from a handful of stars.  Additionally, the potential presence of $s$-process material can complicate interpretations of neutron-capture elements (e.g., \citealt{Roederer2011}).} Measuring the chemical abundances inside the atmospheres of giant metal-poor stars in M15 is therefore essential for determining the progenitor events that occurred during its formation. 

This paper presents analyzes of high-resolution spectra of 62 red giant branch (RGB) and asymptotic giant branch (AGB) stars in M15 to further characterize the $r$-process spreads within the cluster.  The following sections outline steps needed to complete this objective. The targets, observations, and reduced data are mentioned in Section \ref{sec:Obs}. The methods to determine the atmospheric parameters and iron abundances in the sample are described in Section \ref{sec:AtmParmas}. Section \ref{sec:Abunds} then discusses the methods for synthesizing spectral lines of neutron-capture elements, while Section \ref{sec:Discussion} comments on the patterns of the neutron-capture elements, spreads of neutron-capture elements within M15, and the radial distributions of the stars. Finally, Section \ref{sec:Conclusion} concludes this paper with next steps and future work.

\section{Observations}\label{sec:Obs}

Observations were made with the
Michigan/Magellan Fiber System (M2FS)
and MSpec double spectrograph \citep{mateo12,bailey12}
at the Landon Clay (Magellan~II) telescope
at Las Campanas Observatory, Chile.
Our observations used the HiRes gratings and
95~$\mu$m slits on both
spectrographs, and we did not rebin the CCD images.
This setup yielded a spectral resolving power
($R \equiv \lambda/\Delta\lambda$) of 
46,800 on one spectrograph and 32,400 on the other,
as measured from the widths of isolated Th or Ar 
emission lines in the comparison lamp spectra;
the variation is mainly due to the alignment of the 
fiber tetrises relative to the spectrograph
entrance slits.
Order-selecting filters transmitted wavelengths
from $\approx$4425~\AA\ to 
$\approx$4635~\AA\ (orders 77--80) for each fiber.

Three fields were observed over 8~nights
in September, 2017, and August, 2018.
Observations were conducted 
at airmasses ranging from 1.3 to 1.8
and in seeing conditions
that ranged from 0\farcs5 to 1\farcs3.
The total exposure time was 18.3~hr,
divided among the three fields with 
$\approx$5.0 to 6.8~hr per field.
Fibers not assigned to M15 stars were
placed on the sky to facilitate sky subtraction.

We use a custom set of Python routines\footnote{%
\url{https://github.com/baileyji}} to perform the initial image processing,
including subtracting the bias,
merging data from different CCD amplifiers,
stacking images, masking cosmic rays, and
subtracting scattered light.
Image Reduction and Analysis Facility (IRAF)
packages were used for all subsequent data processing,
including flat-fielding, order extraction,
wavelength calibration, spectra co-addition, 
velocity shifting, and continuum normalization.
See \citet{roederer16a} for a more detailed
description of the data reduction and validation
procedures.

A total of 129~stars in M15 were observed.
Targets were selected from the catalogs of
\citet{Carretta2009} and \citet{Kirby2016}.
The brightest red giants were prioritized in our 
fiber assignment process.
A total of 63 of these stars had sufficient
signal-to-noise (S/N) ratios, $> 40$ per pixel
at 4570~\AA,
for abundance analysis.
These 63 metal-poor giants include 
56~RGB stars and 7~AGB stars.   One of the 63 stars was found to be a spectroscopic binary and was excluded from further analysis.

\section{Atmospheric Parameters and Metallicities}\label{sec:AtmParmas}
The 2017 version of the local thermal equilibrium (LTE) line analysis code {\tt MOOG} \citep{Sneden1973}, with an appropriate treatment for scattering \citep{Sobeck2011},\footnote{\url{https://github.com/alexji/moog17scat}} is used to determine chemical abundances.  The spectral lines that are analyzed are shown in Table \ref{table:LineList}.  Additional spectral lines are included in spectrum syntheses, including atomic lines, hyperfine structure (HFS), molecular lines, and isotopic splitting; these line lists are generated with the {\tt linemake} code.\footnote{\vspace{0.5in}\url{https://github.com/vmplacco/linemake}}  Initial equivalent widths (EWs) were measured using the program {\tt DAOSPEC} \citep{Stetson2008}; discrepant lines were then remeasured manually using the {\it splot} tool in the Image Reduction and Analysis Facility program IRAF \citep{Tody1986, Tody1993}.  All abundances are expressed as $\log(\epsilon(X))$ abundances\footnote{$\log{\epsilon\left(X\right)} = \log{\frac{N_{X}}{N_{H}}} + 12$, where X is any element, $N_{X}$ is the column density of element $X$ and $N_{H}$ is the column density of hydrogen H.} or as $[X/Y]$ logarithmic ratios with respect to the Sun.\footnote{For example, $[\text{Fe/H}] = \log \epsilon(\text{Fe}) -  \log \epsilon(\text{Fe})_{\odot}$, where $\log \epsilon(\text{Fe})_{\odot} = 7.50$ \citep{Asplund2009}.}  Unless otherwise noted, the \citet{Asplund2009} solar abundances are adopted.

ATLAS plane-parallel, $\alpha$-enhanced model atmospheres \citep{KuruczModelAtmRef} are used for all stars.  A star's model atmosphere is characterized by the following quantities: the effective temperature, $T_{\text{eff}}$ in K; the surface gravity $\log{g}$ in cgs units; the metallicity $[\rm{M/H}]$; and the microturbulent velocity, $\xi$, in km s$^{-1}$.  The objective is to converge onto an ideal set of atmospheric parameters in order to produce a model atmosphere which is then used as an input to derive abundances.

\startlongtable
\begin{deluxetable}{@{}cccc}
\tabletypesize{\normalsize}
\tablecolumns{8}
\tablewidth{0pt}
\tablecaption{Line list\label{table:LineList}}
\tablehead{
Element & Wavelength (\AA) & EP (eV) & $\log gf$ \\
 }
\decimals
\startdata
\ion{Fe}{1} & 4430.61 & 2.22 & -1.66 \\
\ion{Fe}{1} & 4442.34 & 2.20 & -1.25 \\
\ion{Fe}{1} & 4443.19 & 2.86 & -1.04 \\
\ion{Fe}{1} & 4447.72 & 2.22 & -1.36 \\
\ion{Fe}{1} & 4484.22 & 3.60 & -0.64 \\
\ion{Fe}{1} & 4489.74 & 0.12 & -3.97 \\
\ion{Fe}{1} & 4494.56 & 2.20 & -1.14 \\
\ion{Fe}{1} & 4531.15 & 1.48 & -2.16 \\
\ion{Fe}{1} & 4547.85 & 3.55 & -0.82 \\
\ion{Fe}{1} & 4592.65 & 1.56 & -2.45 \\
\ion{Fe}{1} & 4595.36 & 3.30 & -1.76 \\
\ion{Fe}{1} & 4602.00 & 1.61 & -3.15 \\
\ion{Fe}{1} & 4607.65 & 3.27 & -1.33 \\
\ion{Fe}{1} & 4630.12 & 2.28 & -2.58 \\
\ion{Fe}{2} & 4491.41 & 2.86 & -2.71 \\
\ion{Fe}{2} & 4508.28 & 2.86 & -2.42 \\
\ion{Fe}{2} & 4515.34 & 2.84 & -2.60 \\
\ion{Fe}{2} & 4522.63 & 2.84 & -2.29 \\
\ion{Fe}{2} & 4555.89 & 2.83 & -2.40 \\
\ion{Fe}{2} & 4576.34 & 2.84 & -2.95 \\
\ion{Fe}{2} & 4583.83 & 2.81 & -1.94 \\
\ion{Fe}{2} & 4620.52 & 2.83 & -3.21 \\
\ion{Sr}{1} & 4607.33 & 0.00 &  0.28 \\
\ion{Zr}{2} & 4496.96 & 0.71 & -0.89 \\
\ion{Zr}{2} & 4613.95 & 0.97 & -1.54 \\
\ion{Ba}{2} & 4454.04\tablenotemark{a} & 0.00 & 0.14\\
\ion{La}{2} & 4574.90\tablenotemark{a} & 0.17 & -1.08 \\
\ion{Ce}{2} & 4460.21 & 0.96 & -1.59 \\
\ion{Ce}{2} & 4471.24 & 0.70 &  0.23 \\
\ion{Ce}{2} & 4486.91 & 0.30 & -0.18 \\
\ion{Ce}{2} & 4523.08 & 0.52 & -0.08 \\
\ion{Ce}{2} & 4539.75 & 0.33 & -0.08 \\
\ion{Ce}{2} & 4562.36 & 0.48 &  0.21 \\
\ion{Ce}{2} & 4572.28 & 0.68 &  0.22 \\
\ion{Ce}{2} & 4628.16 & 0.52 &  0.14 \\
\ion{Nd}{2} & 4451.56 & 0.38 &  0.07 \\
\ion{Nd}{2} & 4451.98 & 0.00 & -1.10 \\
\ion{Nd}{2} & 4462.98 & 0.56 &  0.04 \\
\ion{Nd}{2} & 4501.81 & 0.20 & -0.69 \\
\ion{Nd}{2} & 4541.27 & 0.38 & -0.74 \\
\ion{Nd}{2} & 4542.60 & 0.74 & -0.28 \\
\ion{Nd}{2} & 4563.22 & 0.18 & -0.88 \\
\ion{Sm}{2} & 4433.89 & 0.43 & -0.19 \\
\ion{Sm}{2} & 4434.32 & 0.38 & -0.07 \\
\ion{Sm}{2} & 4523.91 & 0.43 & -0.39 \\
\ion{Sm}{2} & 4566.20 & 0.33 & -0.59 \\
\ion{Sm}{2} & 4577.69 & 0.25 & -0.65 \\
\ion{Sm}{2} & 4615.68 & 0.19 & -0.84 \\
\ion{Eu}{2} & 4435.58\tablenotemark{a} & 0.21 & -0.11 \\
\ion{Dy}{2} & 4449.70 & 0.00 & -1.03 \\
\enddata
\medskip
\tablenotetext{a}{This line has hyperfine structure and/or isotopic splitting.}
\end{deluxetable}

\subsection{Atmospheric Parameters}\label{sec:ATMs}

\subsubsection{Initial Photometric Parameters}
An initial set of atmospheric parameters are derived from the photometry of \citet{Stetson1994, Stetson2000} and \citet{Carretta2009}.  Temperatures and initial [Fe/H] ratios were previously derived by \citet{Carretta2009} and \citet{Kirby2016}.  To initialize $\log{g}$, the photometric $T_{\text{eff}}$ and $[\text{Fe/H}]$ are used as inputs to perform an interpolation between two BaSTI isochrones \citep{BaSTIREF,BaSTIREF2}.  A value for $\log{g}$ is obtained from each isochrone using:
\begin{equation}
    \left(\log{g}\right)_{i} = \log{g_{\odot}} + \log{\left(\frac{M}{M_{\odot}}\right)} - \log{\left(\frac{L}{L_{\odot}}\right)} + 4\log{\left(\frac{T_{\text{eff}}}{T_{\text{eff},\odot}}\right)}
\end{equation}
where $i$ corresponds to a specific isochrone and $\log{g}_{\odot} \approx 4.44$ \citep{McWB}.  Note that the values for $M$, $L$, and $T_{\text{eff}}$ are determined by running a routine that searches for the temperature right below the photometric $T_{\text{eff}}$ in the RGB or AGB of the isochrone. From the photometry, M15 targets have a range of $[\text{Fe/H}]$ values that lie between $-2.62$ and $-2.20$. 
The BaSTI isochrones with $[\text{Fe/H}]_{1} = -2.62$ and $[\text{Fe/H}]_{2} = -2.14$ were selected to bracket the photometric M15 values.  The $\log g$ was calculated for each isochrone; the adopted photometric $\log g$ is then a weighted average of the two $\log g$ values, based on the predicted [Fe/H]. This technique is applied to every M15 target.  The photometric surface gravities are then used to derive estimates for the microturbulent velocity, using the empirical relationship derived by \citet{McWB}.  The initial photometric parameters for each star are shown in Table \ref{table:AtmParams}.

\subsubsection{Spectroscopic Parameters}
The final spectroscopic parameters are determined by examining line-to-line trends in the \ion{Fe}{1} abundances as a function of various transition properties, such as the excitation potential (EP, in eV) and the reduced EW (REW = $\log{\left(\text{EW}/\lambda\right)}$).  A negative trend in iron abundance with EP suggests that the temperature in the model atmosphere is too high; this is because the higher temperature model predicts that there will be more electrons in the higher excitation states than there are in reality. Similarly, the parameter that best controls the REW trend is the microturbulent velocity.  Flattening the EP and REW trends therefore leads to spectroscopic values for $T_{\rm{eff}}$ and $\xi$.

One drawback to this method (also known as the excitation method) is that the spectroscopic effective temperature has inevitable systematic offsets from the original photometric temperatures. The final spectroscopic temperature has the tendency to be a few hundred degrees lower than the photometric temperature (e.g., \citealt{Aoki2007, Cayrel2004, Frebel2010, Hollek2011, Johnson2002, Lai2008}).  For that reason, after a spectroscopic $T_{\rm{eff}}$ is identified by flattening trends in iron abundance with EP, the spectroscopic temperature correction of \citet{Frebel2013} is adopted.  Note that the \citet{Frebel2013} correction was determined using metal-poor giants with $-3.3 < [\text{Fe/H}] < -2.5$.  Although many of the M15 stars fall just outside of the calibration region, this correction is still applied for consistency.  A new isochrone-based $\log g$ is then found with this corrected spectroscopic temperature.  Finally, the metallicity of the isochrone is determined from the average [Fe/H] of each star.  

The final spectroscopic parameters for each star are shown in Table \ref{table:AtmParams}.  The random uncertainties in the [\ion{Fe}{1}/H] and [\ion{Fe}{2}/H] abundances are based on the line-to-line dispersion.  For each M15 target, the standard deviation is divided by the square root of the number of lines associated with each Fe atom type. These uncertainties are quoted in Table \ref{table:AtmParams}.

\begin{deluxetable*}{@{}ccccccccccccc}
\tabletypesize{\normalsize}
\tablecolumns{8}
\tablewidth{0pt}
\tablecaption{Photometric and Spectroscopic Parameters for M15 Targets\label{table:AtmParams}}
\tablehead{
 & \multicolumn{2}{l}{Photometric} & \phantom{$-$} & \multicolumn{7}{l}{Spectroscopic$^{a}$} \\
 & $T_{\text{eff}}$ (K) & $\log{g}$   &  & $T_{\text{eff}}$ (K) & $\log{g}$ & $\xi$ (km/s) & $[\text{\ion{Fe}{1}/H}] $ & $N$ & $[\text{\ion{Fe}{2}/H}]$ & $N$ }
\decimals
\startdata
13196 & 4720 & 1.46 & & 4783	&   1.52  &	2.71  &	$-2.57\pm0.06$ & 10 & $-2.61\pm0.12$ & 7 \\
18815 & 4832 & 1.83 & & 4879	&   1.67  &	1.88  &	$-2.42\pm0.09$ & 13 & $-2.50\pm0.03$ & 7 \\
18913 & 4735 & 1.61 & & 4796 & 1.53 & 2.03 & $-2.56\pm0.07$ & 12 & $-2.65\pm0.04$ & 6 \\
21948 & 4746 & 1.66 & & 4820 &  1.60 &	2.29 &  $-2.61\pm0.02$ & 10 & $-2.54\pm0.02$ & 8 \\
2792 &  4567 & 1.26 & & 4510 &  0.83 &	2.11 &  $-2.51\pm0.04$ & 14 & $-2.59\pm0.02$ & 8 \\
28510 & 4754 & 1.64 & & 4679 &  1.25 &	2.06 &  $-2.49\pm0.03$ & 10 & $-2.60\pm0.01$ & 8 \\
28805 & 4836 & 1.65 & & 4900 &  1.79 &  1.01 &  $-2.53\pm0.06$ & 7 & $-2.56\pm0.06$ & 7 \\
31313 & 4805 & 1.75 & & 4724 &  1.42 &	1.89 &  $-2.44\pm0.03$ & 10 & $-2.49\pm0.03$ & 7 \\
31791 & 4978 & 2.13 & & 4957 &  1.85 &	2.28 &  $-2.54\pm0.03$ & 10 & $-2.47\pm0.02$ & 7 \\
\enddata
\medskip
\tablecomments{(This table is available in its entirety in machine-readable form.)}
\end{deluxetable*}

\subsubsection{Uncertainties in Atmospheric Parameters}\label{sec:AtmErrors}
This technique for deriving spectroscopic atmospheric parameters has been used by many other groups.  However, this analysis is complicated by the paucity of available \ion{Fe}{1} lines in the limited spectral range.  Each star has only 7-15 \ion{Fe}{1} lines and 5-8 \ion{Fe}{2} lines, potentially leading to large uncertainties in the atmospheric parameters which can then lead to systematic errors in the abundances. Therefore, it is important to determine the uncertainties of the atmospheric parameters. Two sets of errors in the atmospheric parameters will be calculated using a representative cool star and a representative hot star.

The lower temperature RGB star 37215 is the representative cool star for the sample. In this process, the parameters are changed independently, even though they do have some dependence on each other. The first thing to do is to determine the uncertainty for the temperature (before the correction). Recall that the spectroscopic temperature was found by minimizing trends in EP and $\log \epsilon(\text{\ion{Fe}{1}})$.  The uncertainty in the temperature can then be found from the uncertainty in the slope of the least-squares fit to the points, which leads to an uncertainty in the resulting temperature.  This test yields a value of $\Delta T_{\rm{eff}} = 73 \, \text{K}$.  Secondly, the systematic errors for the surface gravity can be calculated using the uncertainties in the temperature and finding the corresponding isochrone surface gravities.  This produces an uncertainty of $\Delta \left(\log{g}\right)_{\text{sys}} = 0.07$.  Finally, the systematic errors in the microturbulent velocity are determined using the uncertainty in the slope in the REW plot, yielding a value of $\Delta \xi_{\text{sys}} = 0.096 \, \text{km/s}$.

\begin{table}[h!]
\centering
\vspace*{0.1in}
\caption{Uncertainties in Atmospheric Parameters}
\begin{tabular}{ccccccccc}
\hline
Star & $\Delta T_{\text{eff}}$ (K) & $\Delta \log{g}$ & $\Delta \xi$ (km/s) \\
\hline
37215 (Cooler) & 73 & 0.07 & 0.096 \\
36274 (Hotter) & 95 & 0.23 & 0.140 \\
\hline
\end{tabular}
\label{table:errors}
\end{table}

The higher temperature RGB star 36274 is the representative hot star for the sample. The same method as for the cooler star is adopted. The uncertainty in the slope in the trends with EP leads to an uncertainty of $\Delta T_{sys} = 95 \, \text{K}$ in the spectroscopic temperature. The corresponding uncertainty in the isochrone-based surface gravity is $\Delta \left(\log{g}\right)_{\text{sys}} = 0.23$.  Finally, the uncertainty in the microturbulent velocity is found to be $\Delta \xi_{\text{sys}} = 0.14 \, \text{km/s}$.

The final uncertainties in the atmospheric parameters for both representative stars are shown in Table \ref{table:errors}. 


\subsection{Comparisons with Literature Values}\label{subsec:LitCompParams}

\subsubsection{Comparisons with Previous High-Resolution Analyses}\label{subsubsec:CompLit}
Twenty four of the stars in this analysis have been previously observed at medium or high spectral resolution.  \citet{Sneden1997} observed 7 of the target stars at high-resolution in the red; \citet{Sneden2000a} re-observed K341 at high-resolution, further in the blue; \citet{Sneden2000b} analyzed medium-resolution spectra of five additional stars; \citet{Otsuki2006} observed two of these stars at high-resolution; \citet{Sobeck2011} analyzed K341 (including scattering in the analysis for the first time); and \citet{Worley2013} observed eighteen of the target stars at medium resolution.  Some stars overlap between multiple samples.  In general, the high-resolution papers followed a similar procedure as this paper for determining the atmospheric parameters, with the exception of \citet{Sneden2000b}: effective temperatures were determined from minimizing trends in iron abundance with excitation potential, surface gravities were found by forcing the abundances from \ion{Fe}{1} and \ion{Fe}{2} to be equal (an assumption which is known to suffer from non-LTE effects; e.g., \citealt{Kraft2003}), and microturbulent velocities were found by minimizing trends with reduced equivalent width. For the medium-resolution analyses, \citet{Sneden2000b} estimated $T_{\rm{eff}}$ and $\log g$ from photometry and adopted a constant $\xi = 2.0$ km/s for their stars, while \citet{Worley2013} used photometric parameters.

\begin{figure}[ht!]
\begin{center}
\centering
\subfigure{\includegraphics[scale=0.5,trim=0 0.1in 0.45in 0.45in,clip]{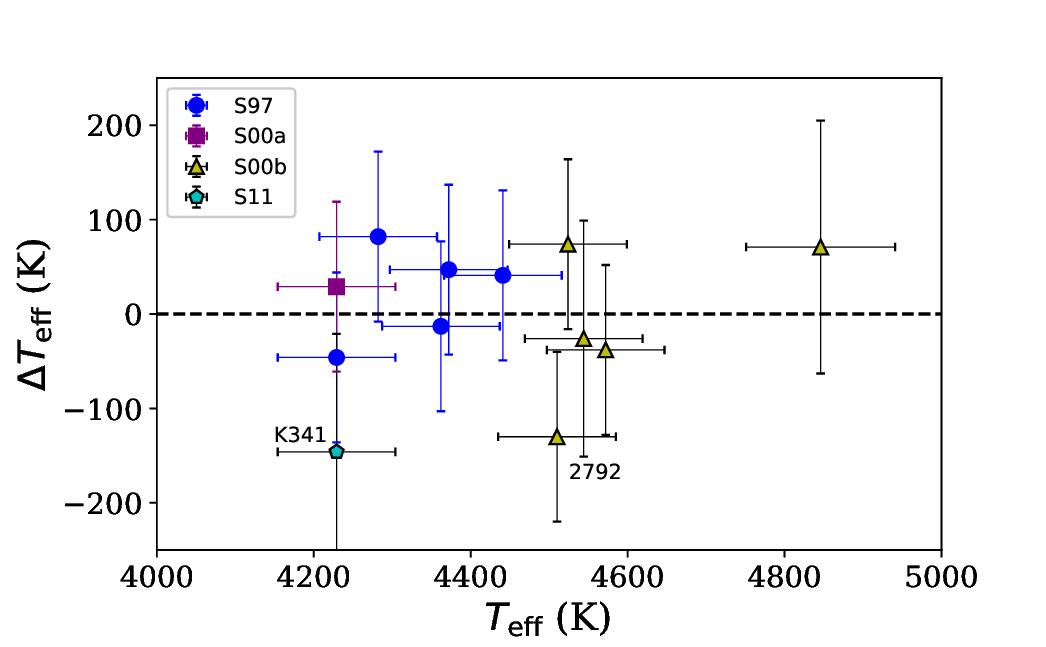}\label{subfig:DeltaTeff}}
\subfigure{\includegraphics[scale=0.5,trim=0 0.1in 0.45in 0.45in,clip]{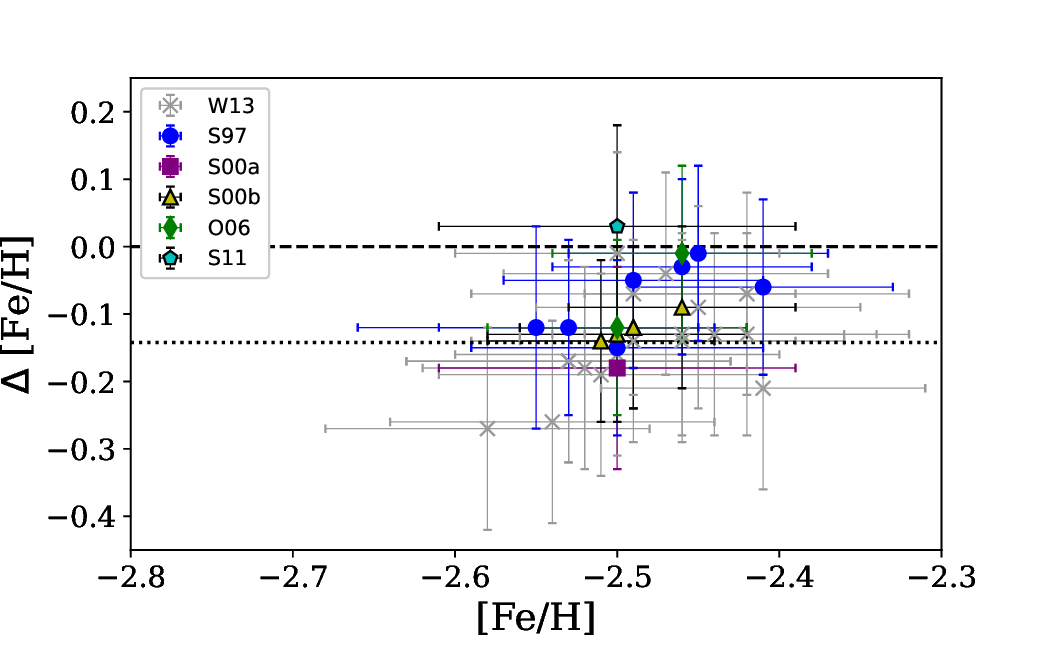}\label{subfig:DeltaFeH}}
\caption{Comparisons between the effective temperature and [Fe/H] from this paper and those in \citet[blue circles]{Sneden1997}, \citet[purple square]{Sneden2000a}, \citet[yellow triangles]{Sneden2000b}, \citet[green diamond]{Otsuki2006}, \citet[cyan pentagon]{Sobeck2011}, and \citet[grey crosses]{Worley2013}.  Note that the \citet{Worley2013} temperatures are not shown in the left panel, since they used photometric temperatures.  Differences are calculated as this work $-$ literature values.\label{fig:LitComp}}
\end{center}
\end{figure}

Figure \ref{fig:LitComp} shows how the effective temperatures and [Fe/H] ratios from this paper compare to these values from the literature.  The temperatures are generally in agreement within the 1$\sigma$ uncertainties; though \citet{Sobeck2011} find a higher temperature for K341, the temperature in this paper agrees with the values from \citet{Sneden1997,Sneden2000a}.  \citet{Sneden2000b} also found a higher temperature for 2792, based on photometry. Although not shown, the surface gravities and microturbulent velocities have predictable offsets given the differences between the analysis methods.  On average the surface gravities in this paper are slightly higher than those achieved from ionization balance and slightly higher than those derived from photometry, though they are generally in agreement within the $1\sigma$ uncertainties.  The microturbulent velocities are also in agreement, with larger offsets for the stars that were observed by \citet[who assumed a constant value for the microturbulent velocity]{Sneden2000b}; however, some of the microturbulent velocities in this work are higher than the values from the literature analyses, such as K146.

\begin{figure}[ht!]
\begin{center}
\centering
\includegraphics[scale=0.75]{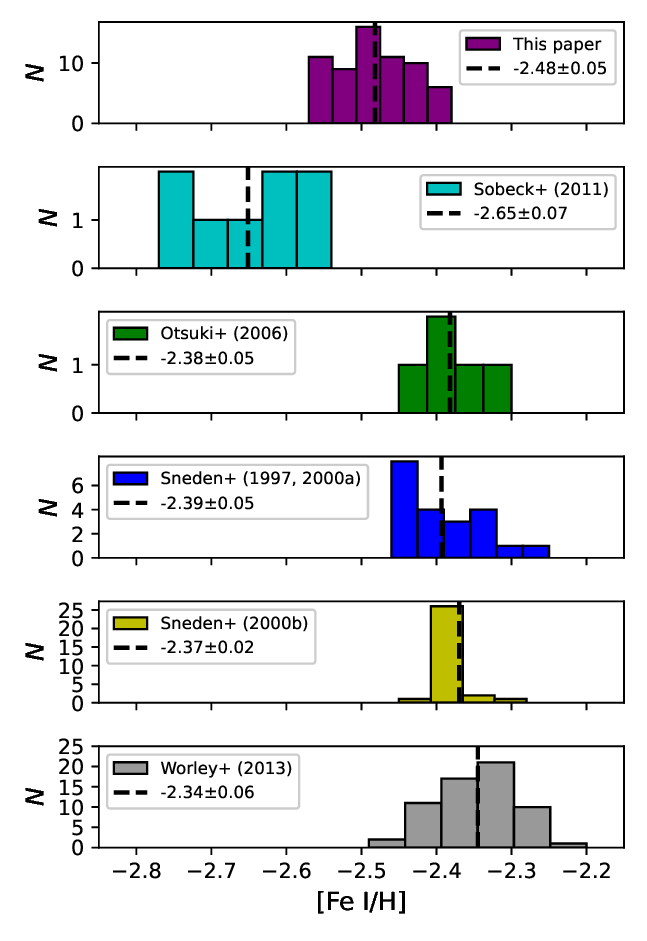}
\caption{Histograms of [\ion{Fe}{1}/H] abundance ratios for all M15 stars, compared to those from the literature \citep{Sneden1997,Sneden2000a,Sneden2000b,Otsuki2006,Sobeck2011,Worley2013}.  \label{fig:FeHHist}}
\end{center}
\end{figure}

Finally, Figure \ref{fig:FeHHist} shows that the [Fe/H] ratios in this paper are generally lower than those from \citet{Sneden1997,Sneden2000a,Sneden2000b,Otsuki2006} and \citet{Worley2013}, but are in agreement with \citet{Sobeck2011}.  This difference is also evident in the distributions of the [\ion{Fe}{1}/H] ratios for all the M15 stars, as seen in Figure \ref{fig:FeHHist}. The mean [Fe/H] lies between the means found in the literature, showing the closest agreement with \citet{Sobeck2011}.  This change in the distribution may be a result of the inclusion of scattering in \texttt{MOOG}, which has the effect of lowering the iron abundances, particularly for spectral lines in the blue \citep{Sobeck2011}.  The offset may also be a result of different solar Fe abundances, though this effect will be small.

\subsubsection{Spectroscopic vs. Photometric Parameters}\label{subsubsec:PhotParams}
\begin{figure}[ht!]
\begin{center}
\centering
\includegraphics[scale=0.55,trim=0.02in 0.1in 0.55in 0.0in,clip]{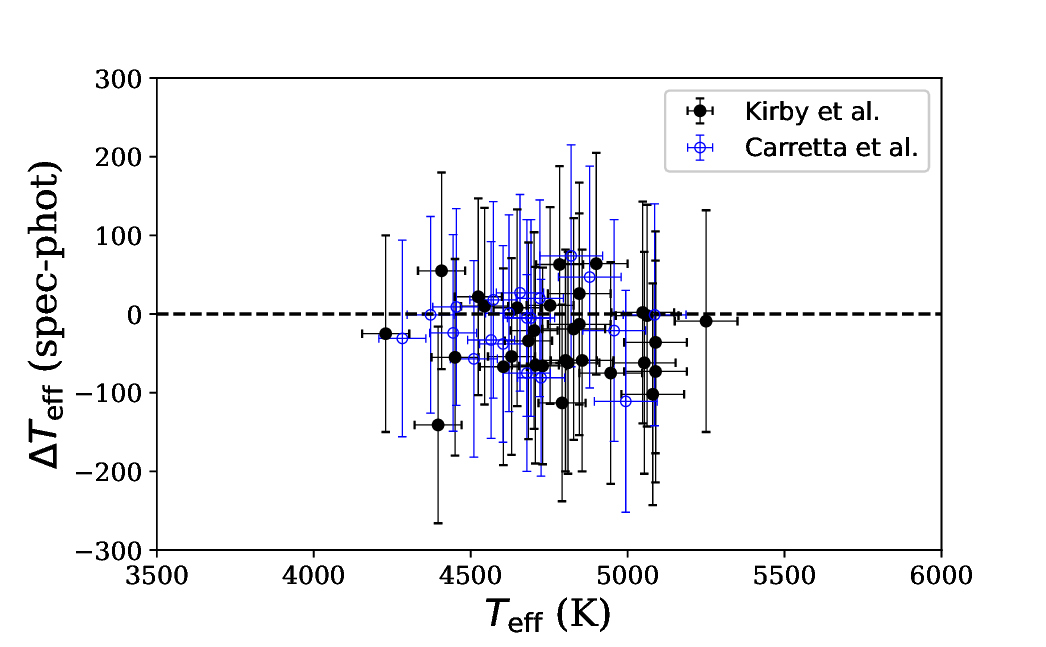}
\caption{A comparison of the spectroscopic and photometric temperatures.\label{fig:Spec_vs_Phot}}
\end{center}
\end{figure}

Figure \ref{fig:Spec_vs_Phot} compares the spectroscopic temperatures with those derived from photometry.  The spectroscopic temperatures are generally in agreement with the photometric temperatures, though on average the spectroscopic temperatures are slightly lower.  This trend is typical for spectroscopic temperatures; although the \citet{Frebel2013} correction has been applied, it may not be sufficient to bring the spectroscopic temperatures up to the photometric values.

\section{Neutron Capture Abundances}\label{sec:Abunds}

Once the atmospheric parameters have been determined for each M15 target, the abundances of the neutron-capture elements can be determined via spectrum syntheses.  \new{For each synthesis, continuum levels were identified by synthesizing the $\sim10$ \AA$\;$ region around the line of interest and minimizing the residuals between the observed and synthetic spectra.}

\subsection{Differential Analysis Techniques}\label{subsec:DiffAnal}
Because of the small numbers of spectral lines available for the analysis, elements with more than one line can have large line-to-line spreads that are the result of variations in atomic data.  To ameliorate this effect, one of the stars, 2792, is adopted as a standard star.  Star 2792 is one of the higher S/N targets in this sample, and has robust measurements of every spectral line used in this analysis.  For that reason, all [X/H] abundances for neutron-capture lines are calculated line-by-line, relative to Star 2792.  The average offsets for each star are then applied to the average abundance in 2792.

\subsection{Barium and Europium}\label{subsec:BaEu}

\begin{figure*}[ht!]
\begin{center}
\centering
\hspace*{-0.45in}
\includegraphics[scale=0.8]{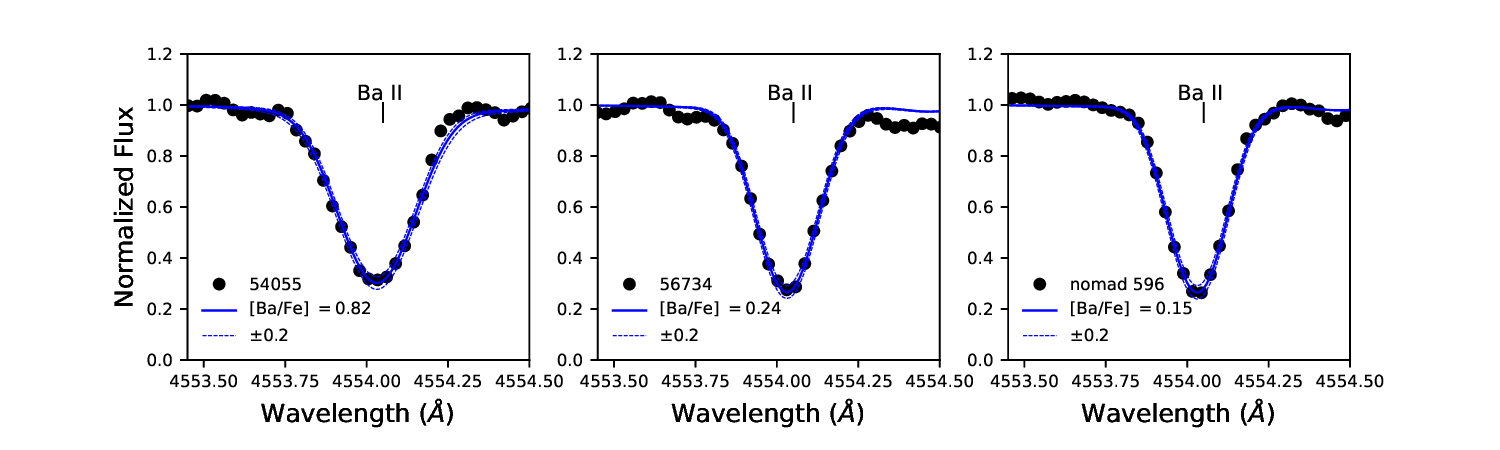}
\caption{Syntheses of the 4554 \AA \hspace{0.015in} \ion{Ba}{2} line in three M15 stars. The observed spectra are shown with black points, while the best fit synthetic spectra are shown with a solid blue line.  The dashed blue lines correspond to variations of $\pm0.2$ dex in the Ba abundance.  Note that for most of these stars the uncertainties are less than 0.2 dex, but these offsets are not easily visible in the plot.\label{fig:BaSynths}}
\end{center}
\end{figure*}

\begin{figure*}[ht!]
\begin{center}
\centering
\hspace*{-0.45in}
\includegraphics[scale=0.8]{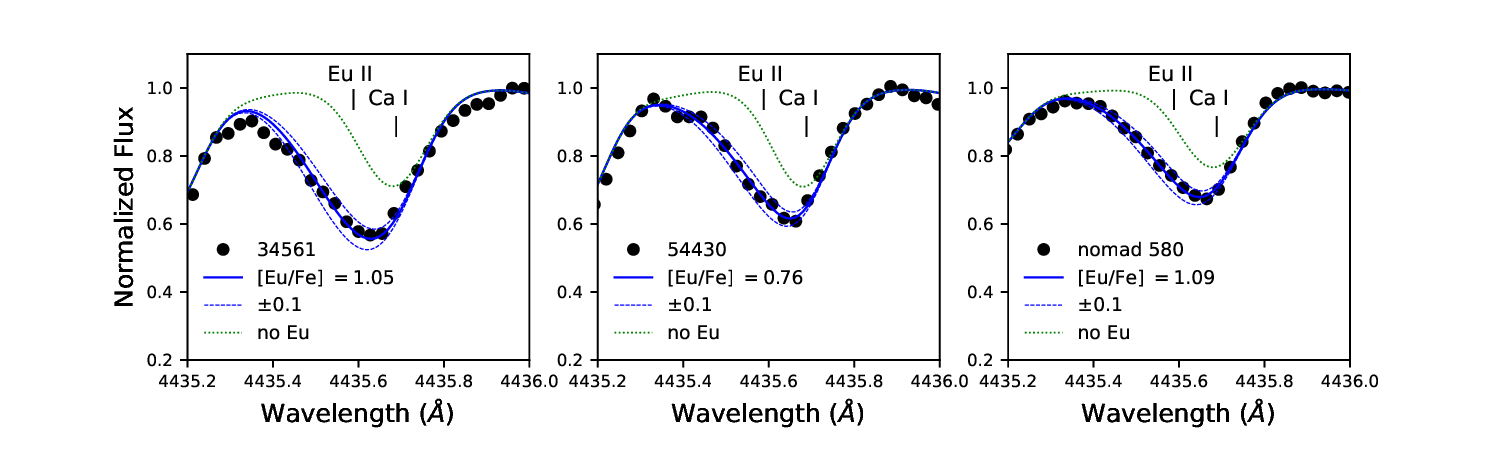}
\caption{Syntheses of the 4435 \AA \hspace{0.015in} \ion{Eu}{2} line in three M15 stars. The observed spectra are shown with black points, while the best fit synthetic spectra are shown with a solid blue line.  The dashed blue lines correspond to variations of $\pm0.1$ dex in the Eu abundance.  The nearby \ion{Ca}{1} line is also identified, for reference. 
 \new{The dotted green line shows a synthesis with no Eu, isolating the Ca line.}\label{fig:EuSynths}}
\end{center}
\end{figure*}

One line each for \ion{Ba}{2} and \ion{Eu}{2} is available in this spectral range.  The \ion{Ba}{2} 4554~\AA\hspace{0.015in} line is quite strong, and has significant isotopic shifts that must be included.  The \ion{Eu}{2} 4435~\AA\hspace{0.015in} line is a blend with the \ion{Ca}{1} 4435~\AA\hspace{0.015in} line, so the synthesis analysis needs to be treated carefully. \new{M15 is known to be a Ca-enhanced cluster, with stellar Ca abundances ranging from $[\rm{Ca/Fe}] = +0.1$ to $+0.5$ \citep{Sneden1997,Sobeck2011}.  For this paper, a value of $[\rm{Ca/Fe}] = +0.3$ was chosen by default.  The Ca abundance was allowed to vary by $\pm0.1$ dex to fit a \ion{Ca}{1} line at 4454.78 \AA$\;$ (which is blended with a weak \ion{Sm}{2} feature).  For a highly $r$-process enhanced star, like 2792, the uncertain Ca abundance can lead to a $\la 0.05$ uncertainty in the Eu abundance; this uncertainty could increase to $\la 0.10$ for the stars with the lowest Eu abundance.} Sample syntheses of the Ba and Eu lines are shown in Figures \ref{fig:BaSynths} and \ref{fig:EuSynths}, respectively.  Uncertainties in the abundances are determined based on the upper and lower limits of the fits.  The [Ba/Fe] and [Eu/Fe] ratios were calculated with respect to \ion{Fe}{2}, and are shown, along with [Ba/Eu] ratios, in Table \ref{table:BaEu}.

\begin{figure}[h!]
\begin{center}
\centering
\hspace*{-0.15in}
\subfigure{\includegraphics[scale=0.5,trim=0.in 0.05in 0.5in 0.45in,clip]{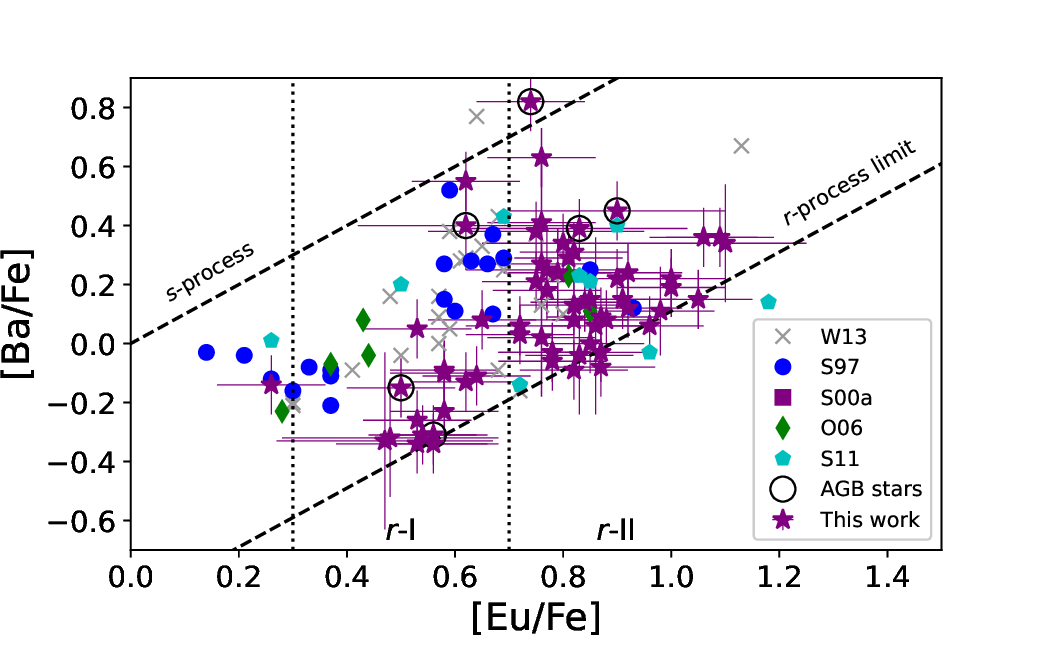}}
\subfigure{\includegraphics[scale=0.5,trim=0.2in 0.05in 0.5in 0.45in,clip]{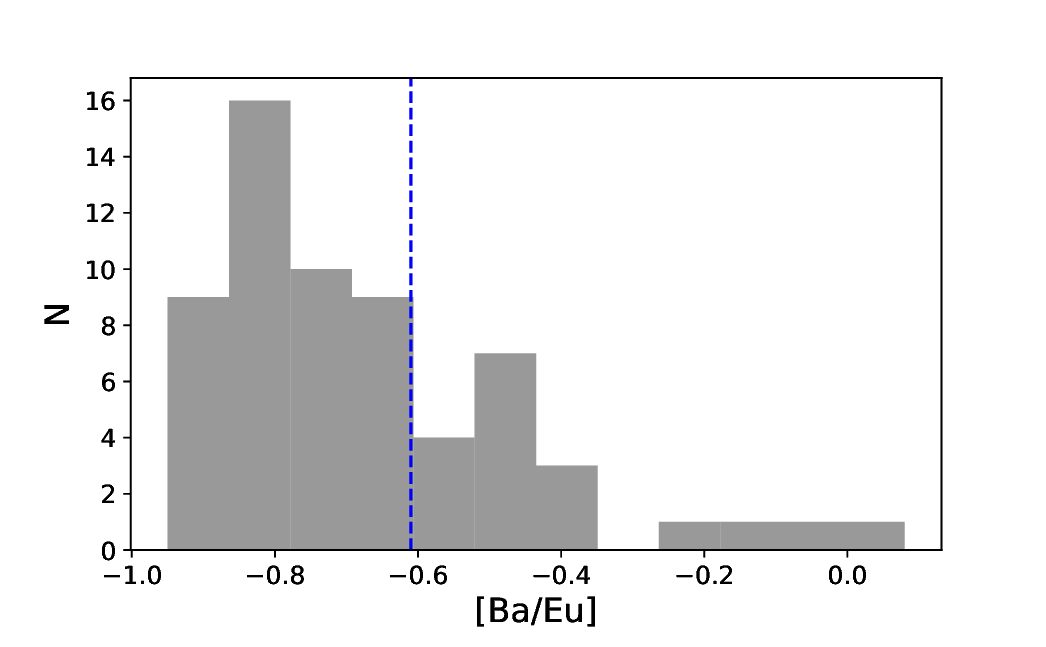}\label{fig:BaEu_dist}}
\caption{{\it Top: } [Ba/Fe] vs. [Eu/Fe] ratios for the 62 M15 stars in this analysis, compared with the literature.  The dashed slanted lines show the traditional boundaries for the $r$-process and the $s$-process.  The $r$-process limit is set to $[\rm{Ba/Eu}] = -0.89$ \citep{Burris2000}, while the boundary between the $r$- and $s$-processes is set at $[\rm{Ba/Eu}] = 0$.  The vertical dotted lines show the boundaries for $r$-I ($+0.3\ge$[Eu/Fe]$<+0.7$) and $r$-II ([Eu/Fe]$\ge+0.7$) stars \citep{Holmbeck2020}.  AGB stars are identified with circles---note that the sole $s$-process-enhanced star in this sample is an AGB star.  {\it Bottom: } The overall distribution of [Ba/Eu] ratios in the sample.  The majority of the stars appear to have an $r$-process signature, while some have signs of enhanced Ba.  The dashed blue line shows the adopted distinction between stars with high and low Ba (see the text).
\label{fig:BaEu}}
\end{center}
\end{figure}

Table \ref{table:BaEu} also shows the classifications of the stars according to their $r$-process enhancement \citep{BeersChristlieb2005,Holmbeck2020}, where $r$-I stars are moderately enhanced in the $r$-process ($+0.3\ge$[Eu/Fe]$<+0.7$) and $r$-II stars are highly enhanced in the $r$-process ([Eu/Fe]$>+0.7$).  The $r$-I and $r$-II classifications also  require that $[\rm{Ba/Eu}] < 0$, a general criterion to exclude stars that are enhanced in the $s$-process. These separation between the $r$- and $s$-processes can be seen in Figure \ref{fig:BaEu}, which also shows stars from the literature, for reference.

\begin{deluxetable*}{@{}lDDDc}
\newcolumntype{d}[1]{D{,}{\;\pm\;}{#1}}
\tabletypesize{\normalsize}
\tablecolumns{8}
\tablewidth{0pt}
\tablecaption{Ba and Eu abundance ratios with uncertainties and $r$-process classifications\label{table:BaEu}}
\tablehead{
Star & \multicolumn{2}{c}{[Ba/Fe]} & \multicolumn{2}{c}{[Eu/Fe]} & \multicolumn{2}{c}{[Ba/Eu]} & Classification}
\decimals
\startdata
13196 & -0.33\pm0.3 & 0.47\pm0.2 & -0.8\pm0.36 & r-I \\
18815 & -0.32\pm0.2 & 0.48\pm0.2 & -0.8\pm0.28 & r-I \\
18913 & -0.31\pm0.1 & 0.54\pm0.1 & -0.85\pm0.14 & r-I \\
21948 & 0.05\pm0.1 & 0.53\pm0.1 & -0.48\pm0.14 & r-I \\
2792 & 0.08\pm0.1 & 0.88\pm0.1 & -0.8\pm0.14 & r-II \\
28510 & -0.03\pm0.1 & 0.78\pm0.1 & -0.81\pm0.14 & r-II \\
31313 & 0.13\pm0.1 & 0.82\pm0.1 & -0.69\pm0.14 & r-II \\
31791 & -0.03\pm0.1 & 0.87\pm0.1 & -0.9\pm0.14 & r-II \\
31914 & 0.36\pm0.1 & 1.06\pm0.1 & -0.7\pm0.14 & r-II \\
\enddata
\medskip
\tablecomments{(This table is available in its entirety in machine-readable form.)}
\end{deluxetable*}

\begin{figure}[h!]
\begin{center}
\centering
\subfigure{\includegraphics[scale=0.5,trim=0 0.1in 0.45in 0.45in,clip]{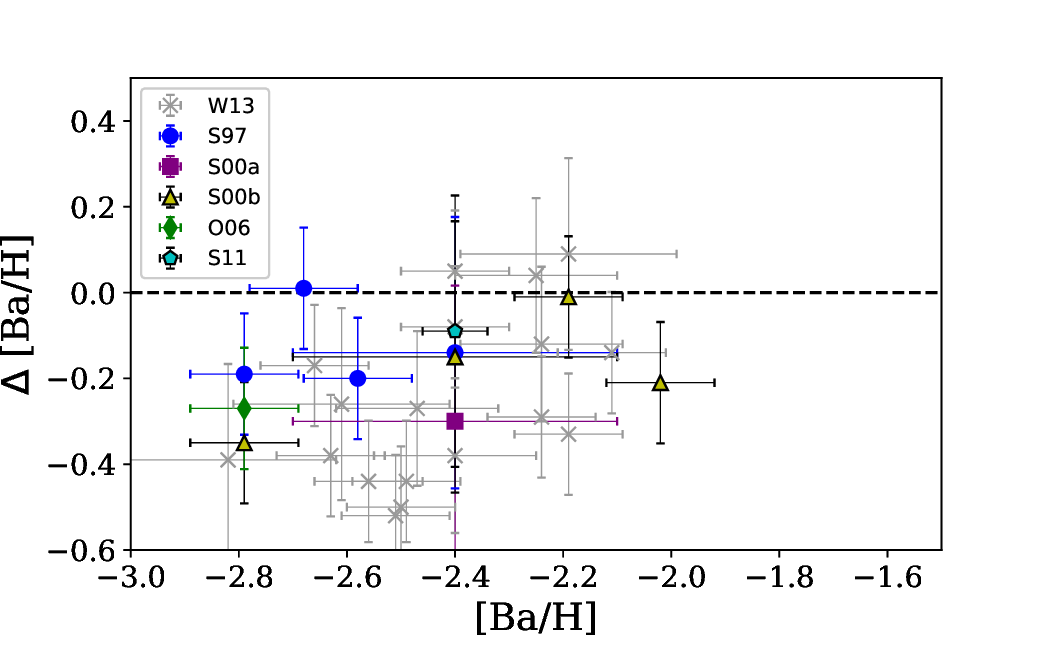}\label{subfig:DeltaBaH}}
\subfigure{\includegraphics[scale=0.5,trim=0 0.1in 0.45in 0.45in,clip]{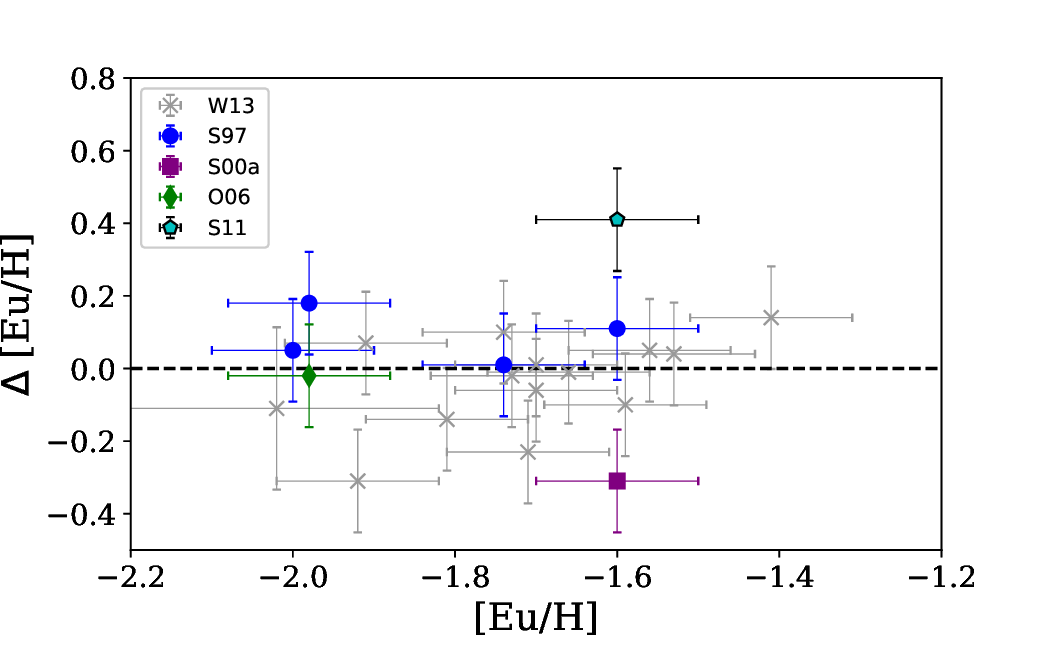}\label{subfig:DeltaEuH}}
\caption{Comparisons between the [Ba/H] ratios (left) and the [Eu/H] ratios (right) from this paper and those in \citet[blue circles]{Sneden1997}, \citet[purple square]{Sneden2000a}, \citet[yellow triangles]{Sneden2000b}, \citet[green diamond]{Otsuki2006}, \citet[cyan pentagon]{Sobeck2011}, and \citet[grey crosses]{Worley2013}.
\label{fig:CompNC}}
\end{center}
\end{figure}

Figure \ref{fig:CompNC} shows how the [Ba/H] and [Eu/H] abundances in this paper compare with those from the literature. The [Eu/H] abundances are generally in good agreement, with an average offset of $\Delta[\rm{Eu/H}] = -0.01\pm0.16$.  The [Ba/H] abundances vary more significantly, with an average offset of $\Delta[\rm{Ba/H}] = -0.22\pm0.16$ and a hint of a possible trend.  The offsets in [Ba/H] may be related to the reliability of the strong 4554 \AA \hspace{0.02in} line, as will be discussed in Section \ref{subsec:Patterns}.

\subsection{Strontium and Zirconium}\label{subsec:SrZr}
In the observed wavelength range, there is one \ion{Sr}{1} line available (at 4607.33 \AA) and two \ion{Zr}{2} lines (at 4496.96 and 4613.95 \AA).  The \ion{Sr}{1} is very weak and is only detectable in 11 stars.  Upper limits on the Sr abundance can, in some cases, at least rule out the presence of strong enhancement in the light neutron-capture elements.  The 4496.96 \AA \hspace{0.02in} \ion{Zr}{2} line was detectable in nearly every star, while the 4613.95 \AA \hspace{0.02in} line was only detectable in 9 stars.  

\subsection{Other Lanthanides}\label{subsec:Lanthanides}
There are lines for 5 other elements in the observed spectra: \ion{La}{2} (1 line), \ion{Ce}{2} (8 lines), \ion{Nd}{2} (7 lines), \ion{Sm}{2} (7 lines), and \ion{Dy}{2} (1 line).  Lines with significant isotopic splitting were not included.  As discussed in Section \ref{subsec:DiffAnal}, the average [X/H] ratios for elements with more than one line were calculated relative to Star 2792.  Not all lines were detectable in all stars.  Stars with lower Eu abundances often showed no detectable lines for the other lanthanides, particularly in lower S/N spectra.  Some lines, such as the \ion{La}{2} line, were occasionally too weak to detect above the noise.  The \ion{La}{2} line is also a blend with an \ion{Fe}{1} line, which sometimes made the \ion{La}{2} line difficult to detect.  

The abundances of the lanthanides are given in Table \ref{table:NCAbunds}.  The quoted uncertainties are the uncertainties in the average abundances.

\begin{deluxetable*}{@{}lDDDDDDD}
\newcolumntype{d}[1]{D{,}{\;\pm\;}{#1}}
\tabletypesize{\normalsize}
\tablecolumns{8}
\tablewidth{0pt}
\tablecaption{Abundances of other neutron capture elements\label{table:NCAbunds}}
\tablehead{
Star & \multicolumn{2}{c}{[Sr/H]} & \multicolumn{2}{c}{[Zr/H]} & \multicolumn{2}{c}{[La/H]} & \multicolumn{2}{c}{[Ce/H]} & \multicolumn{2}{c}{[Nd/H]} & \multicolumn{2}{c}{[Sm/H]} & \multicolumn{2}{c}{[Dy/H]}}
\decimals
\startdata
13196 & \multicolumn{2}{c}{$<$-2.16} & \multicolumn{2}{c}{$<$-1.86} & \multicolumn{2}{c}{---} & \multicolumn{2}{c}{---} & \multicolumn{2}{c}{---} & \multicolumn{2}{c}{---} & \multicolumn{2}{c}{---} \\
18815 & \multicolumn{2}{c}{$<$-2.22} & -2.09\pm0.10 & \multicolumn{2}{c}{---} & \multicolumn{2}{c}{---} & \multicolumn{2}{c}{---} & \multicolumn{2}{c}{---} & \multicolumn{2}{c}{---} \\
18913 & \multicolumn{2}{c}{$<$-2.46} & -2.25\pm0.10 & \multicolumn{2}{c}{---} & \multicolumn{2}{c}{---} & \multicolumn{2}{c}{---} & \multicolumn{2}{c}{---} & \multicolumn{2}{c}{---} \\
21948 & \multicolumn{2}{c}{$<$-2.50} & -2.20\pm0.10 & \multicolumn{2}{c}{---} & \multicolumn{2}{c}{---} & \multicolumn{2}{c}{---} & \multicolumn{2}{c}{---} & \multicolumn{2}{c}{---} \\
2792 & -2.68\pm0.10 & -2.15\pm0.10 & -1.93\pm0.10 & -2.21\pm0.06 & -1.92\pm0.09 & -1.78\pm0.06 & -1.56\pm0.10 \\
\enddata
\tablecomments{(This table is available in its entirety in machine-readable form.)}
\end{deluxetable*}

\section{Discussion}\label{sec:Discussion}

\subsection{Patterns of neutron capture elements}\label{subsec:Patterns}
\begin{figure}[h!]
\begin{center}
\centering
\hspace*{-0.15in}
\includegraphics[scale=0.55,trim=0.in 0.1in 0.5in 0.2in,clip]{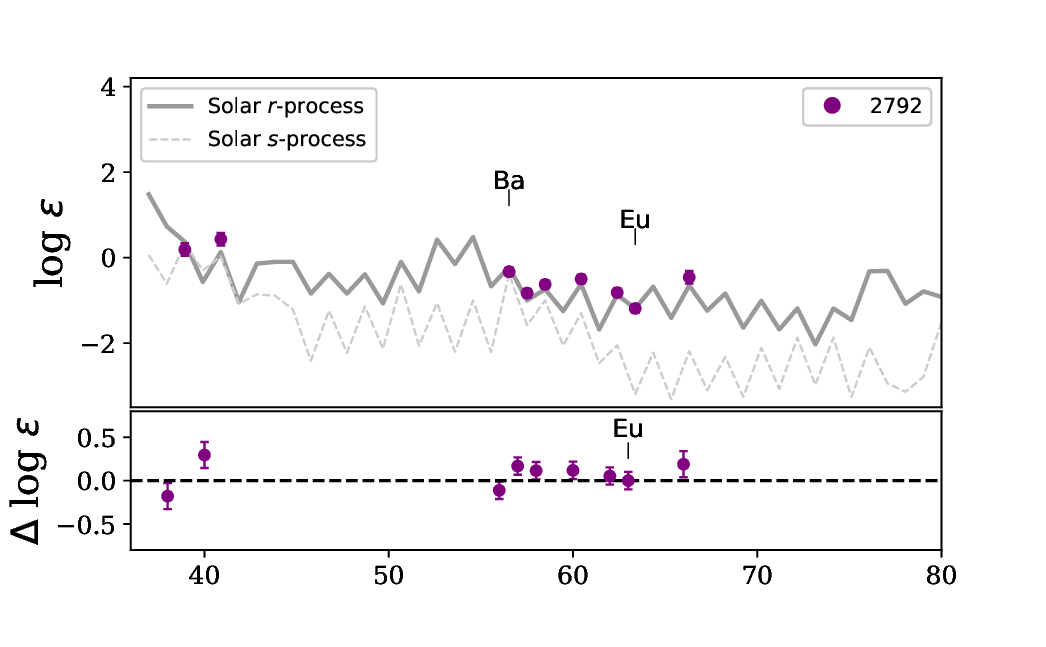}
\caption{The neutron-capture abundances of 2792, compared to the Solar $s$- and $r$-process patterns (from \citealt{Arlandini1999}). The solar patterns are shifted to match the Eu abundance (for the $r$-process) and the Ba abundance (for the $s$-process).\label{fig:2792Pattern}}
\end{center}
\end{figure}

The abundance pattern of the neutron-capture elements in Star 2792 is shown in Figure \ref{fig:2792Pattern}, compared with the Solar $s$- and $r$-process patterns.  The bottom panel of Figure \ref{fig:2792Pattern} shows that star 2792 is generally consistent with the Solar $r$-process pattern, though the Zr and Dy abundances are slightly higher than the Solar pattern and the Ba abundance is slightly lower.  The difficulties of determining the Ba abundance from the strong 4554 \AA \hspace{0.02in} line will be discussed further below.

\begin{figure*}[h!]
\begin{center}
\centering
\subfigure{\includegraphics[scale=0.55,trim=0.in 0.1in 1.2in 0.2in,clip]{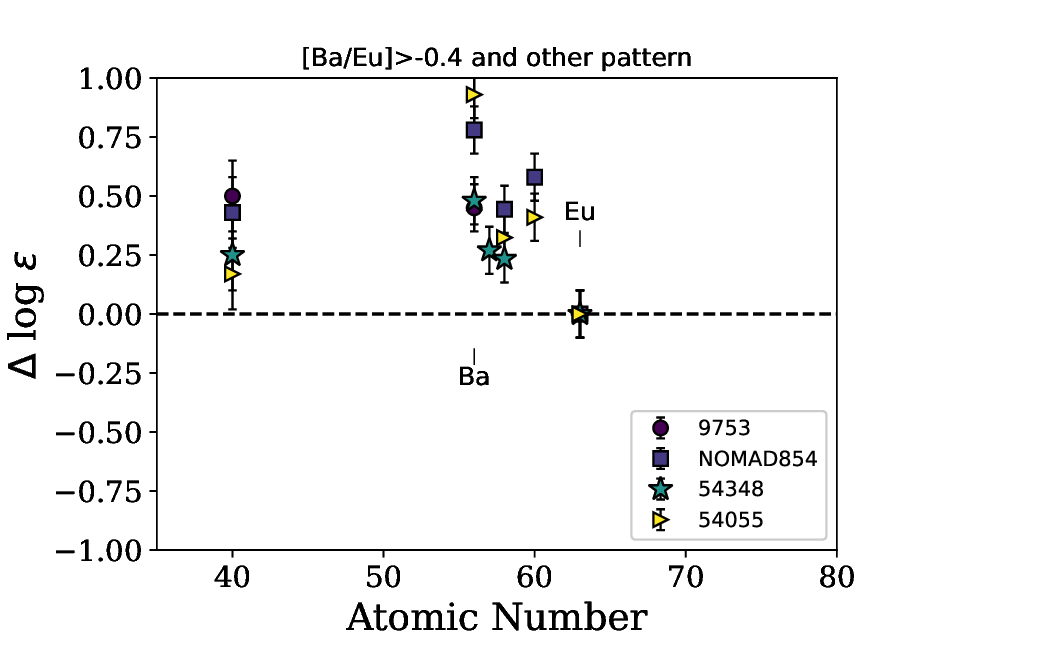}}
\subfigure{\includegraphics[scale=0.55,trim=0.in 0.1in 1.2in 0.2in,clip]{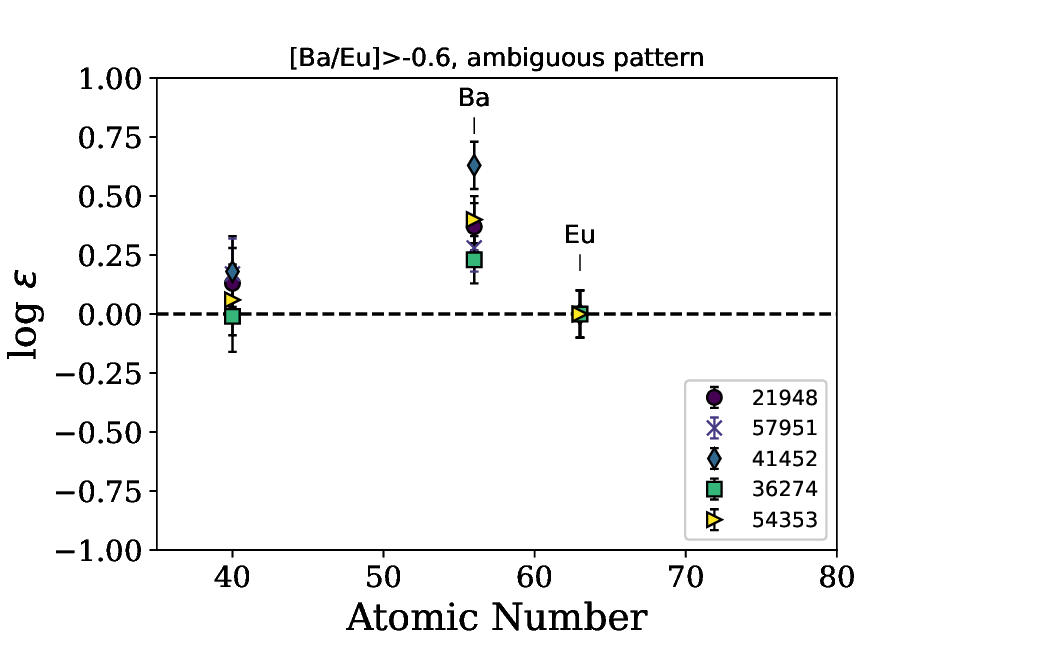}}
\subfigure{\includegraphics[scale=0.55,trim=0.in 0.1in 1.2in 0.2in,clip]{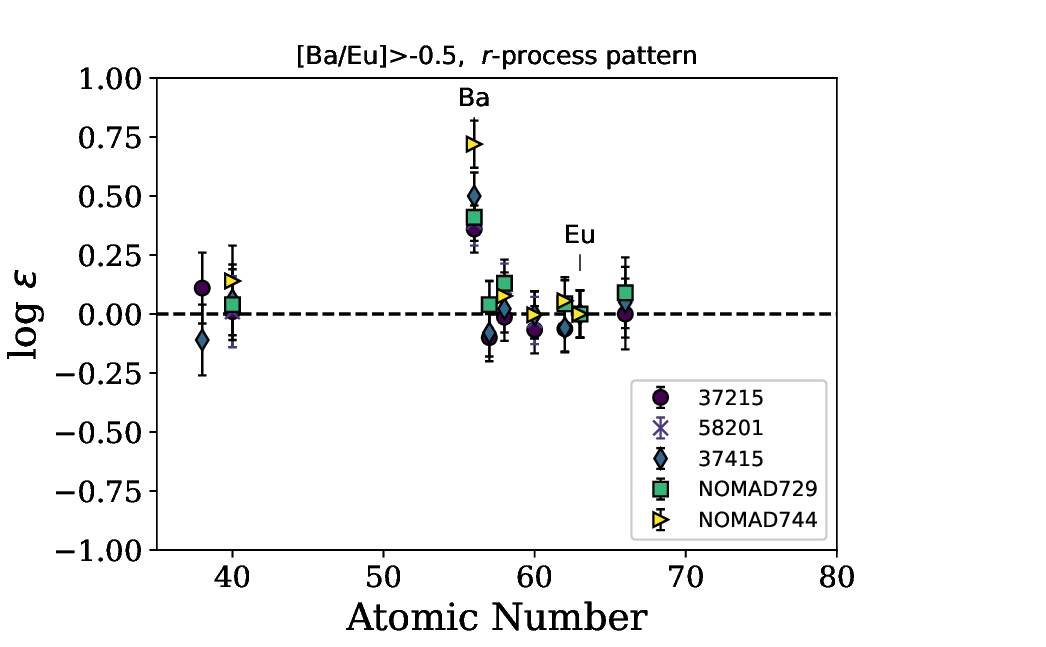}}
\subfigure{\includegraphics[scale=0.55,trim=0.in 0.1in 1.2in 0.2in,clip]{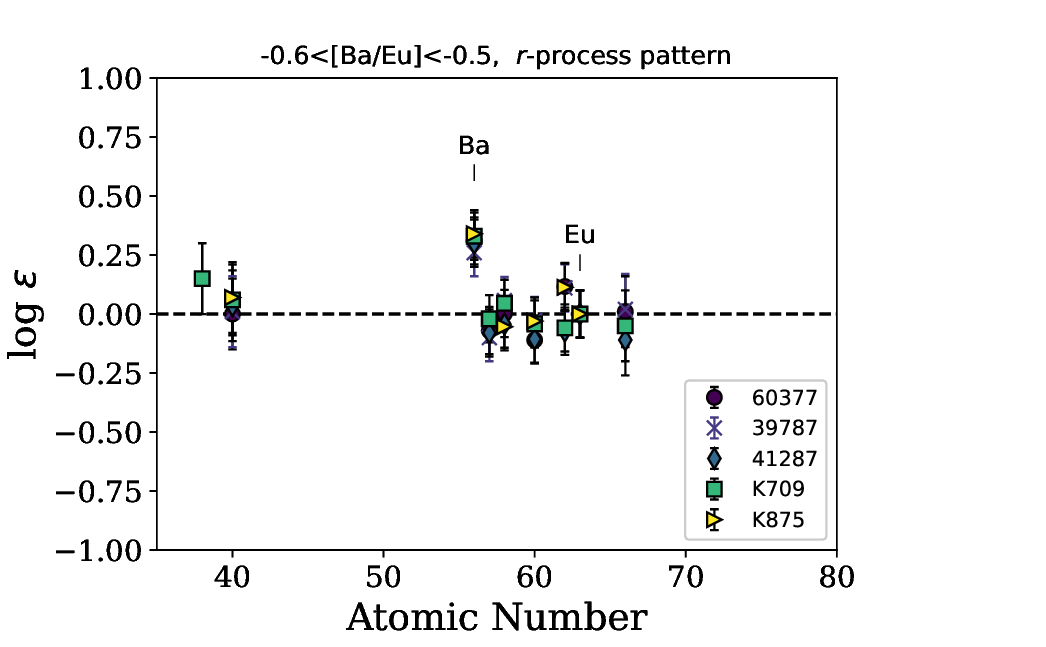}}
\caption{Patterns of neutron-capture elements in the target stars with $[\rm{Ba/Eu}]>-0.6$.  All stars are shifted to a common Eu abundance.  The top left panel shows stars with signs of $s$-process contamination, based on high Ba, La, Ce, etc. The top right panel shows stars with high [Ba/Eu], but an ambiguous pattern.  The bottom two panels show stars with $[\rm{Ba/Eu}]>-0.5$ (left) and $-0.6<[\rm{Ba/Eu}]<-0.5$ (right) but where all elements other than Ba are consistent with the pattern in 2792.
 \label{fig:Patterns_s}}
\end{center}
\end{figure*}

\begin{figure*}[h!]
\begin{center}
\centering
\subfigure{\includegraphics[scale=0.55,trim=0.in 0.1in 1.2in 0.2in,clip]{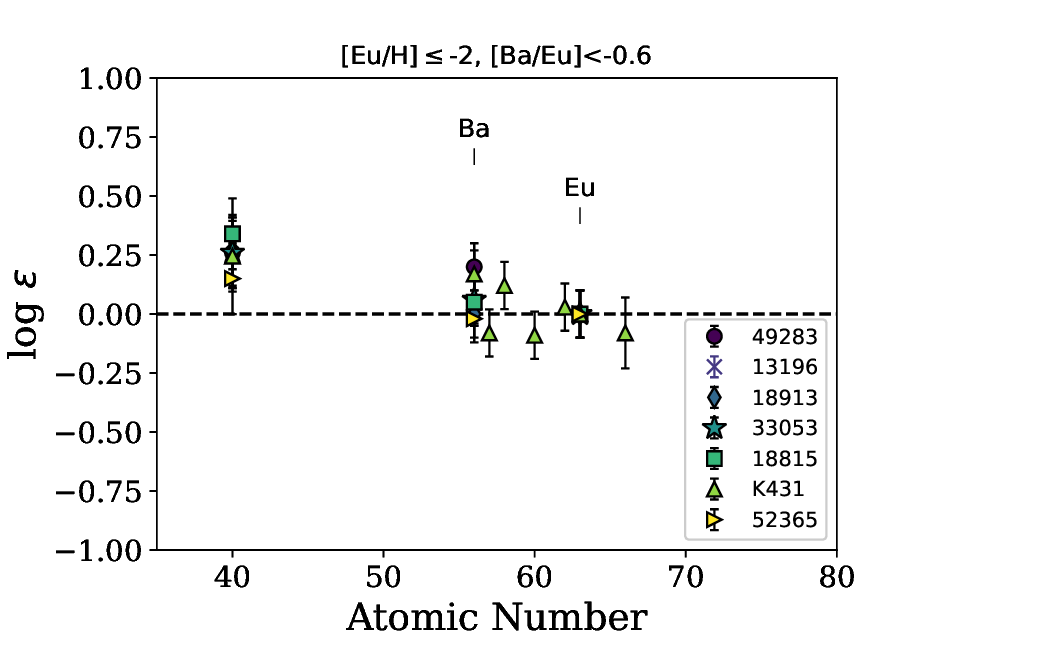}}
\subfigure{\includegraphics[scale=0.55,trim=0.in 0.1in 1.2in 0.2in,clip]{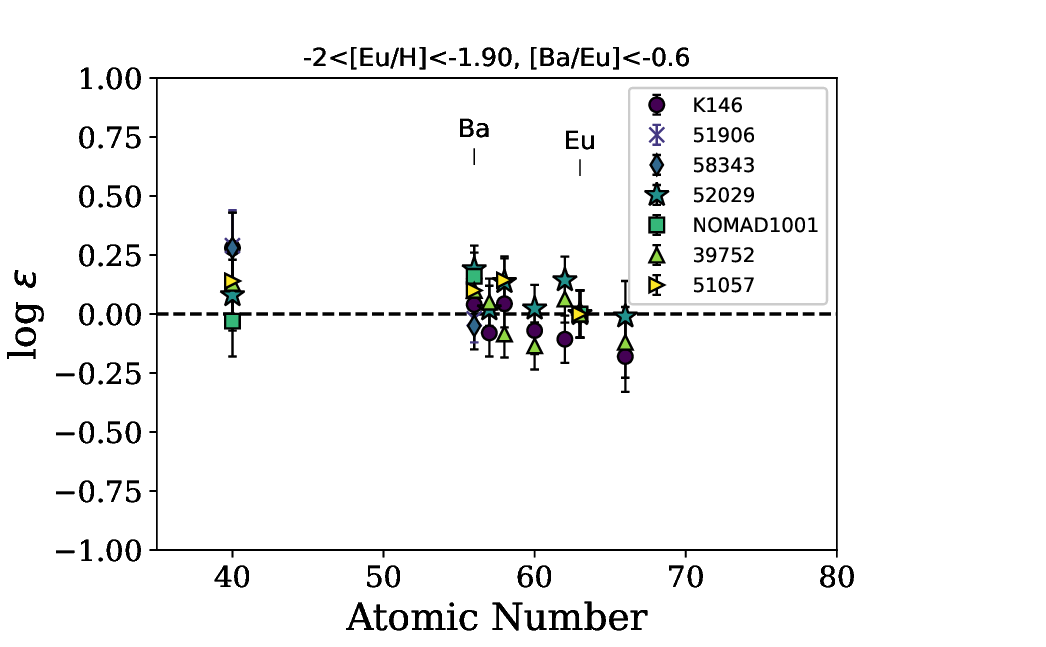}}
\subfigure{\includegraphics[scale=0.55,trim=0.in 0.1in 1.2in 0.2in,clip]{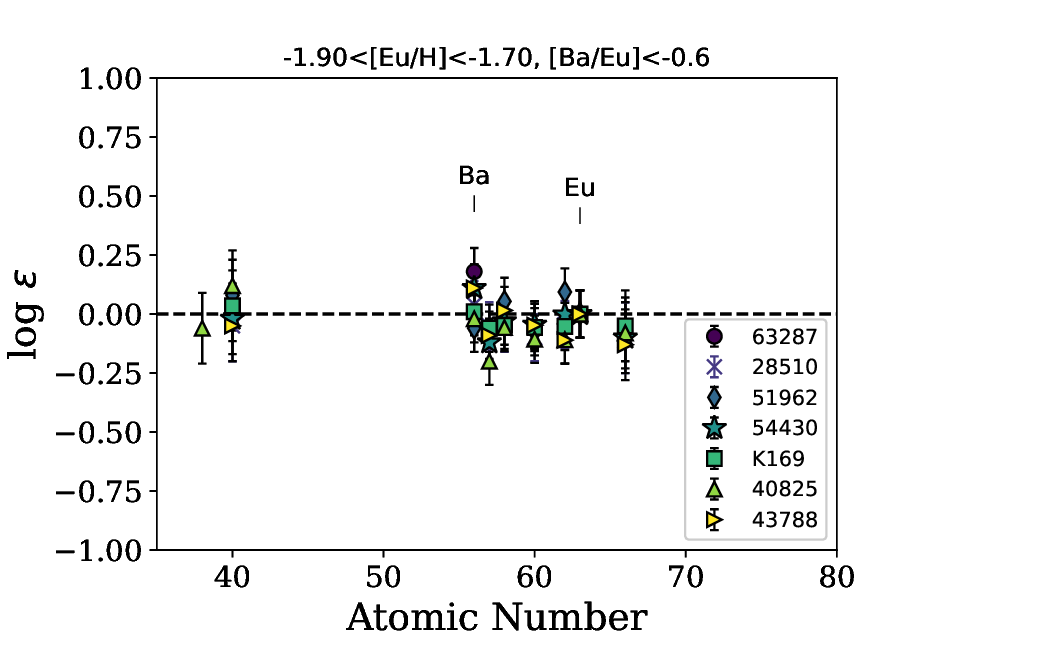}}
\subfigure{\includegraphics[scale=0.55,trim=0.in 0.1in 1.2in 0.2in,clip]{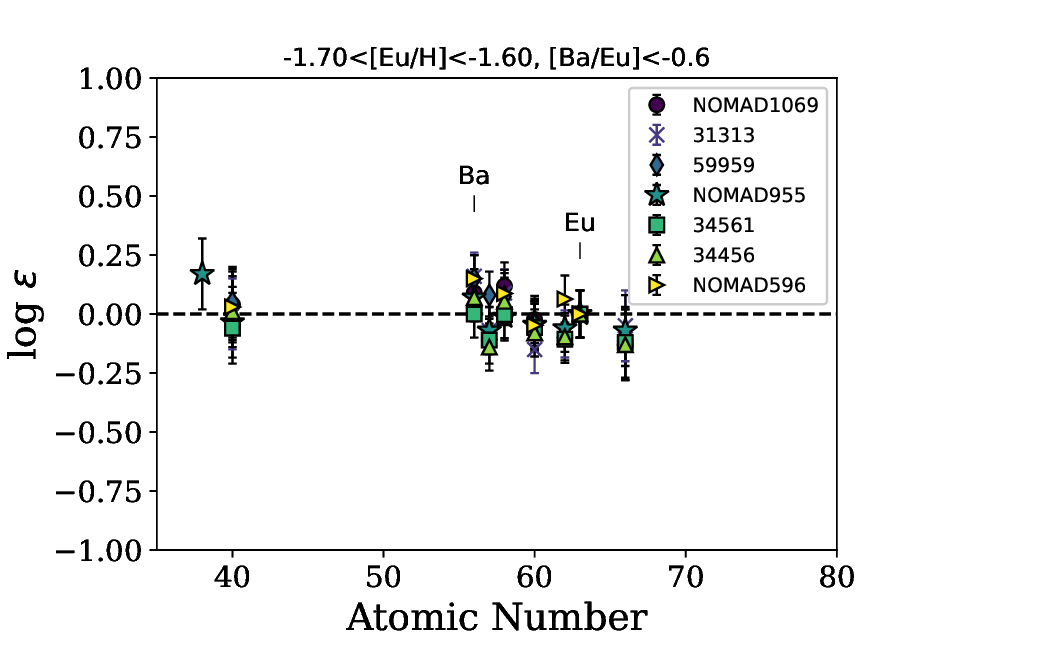}}
\subfigure{\includegraphics[scale=0.55,trim=0.in 0.1in 1.2in 0.2in,clip]{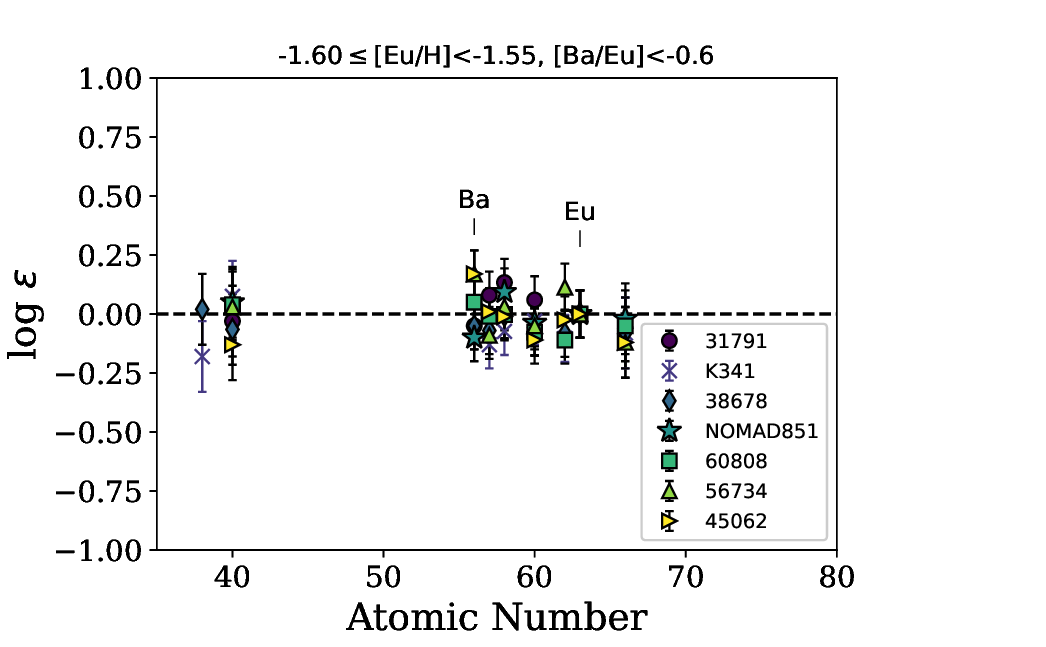}}
\subfigure{\includegraphics[scale=0.55,trim=0.in 0.1in 1.2in 0.2in,clip]{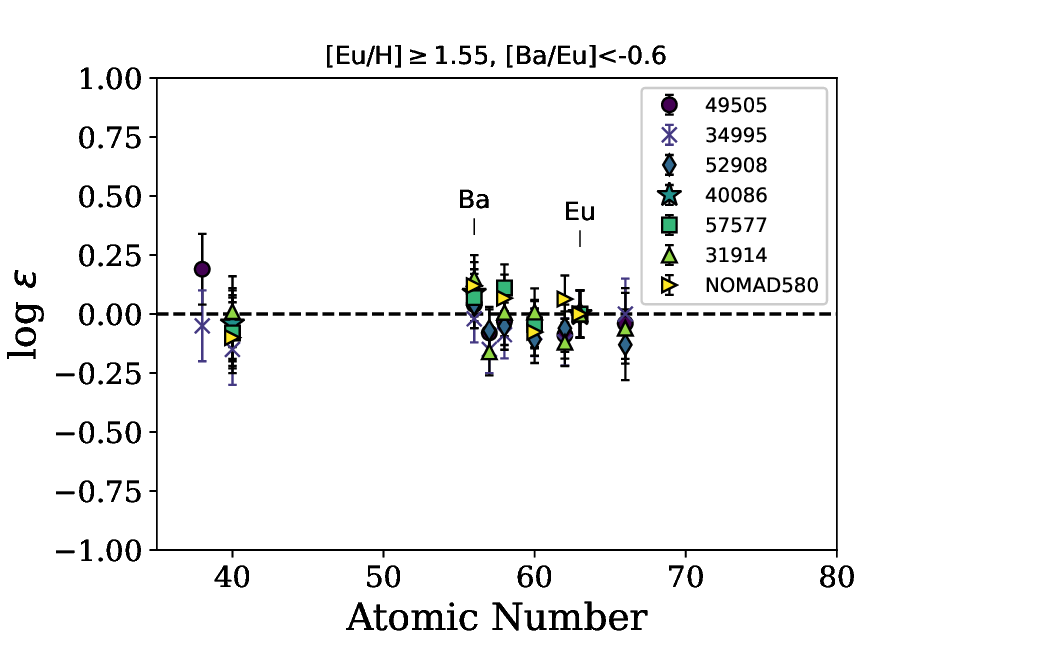}}
\caption{Patterns of neutron-capture elements in the target stars with $[\rm{Ba/Eu}]<-0.6$. All stars are shifted to a common Eu abundance.  
 The panels are arranged from top left to bottom right by increasing [Eu/H].\label{fig:Patterns_r}}
\end{center}
\end{figure*}

Figures \ref{fig:Patterns_s} and \ref{fig:Patterns_r} then show the patterns of the neutron-capture elements in the other M15 stars, relative to the pattern in Star 2792.  The stars have been grouped based on their [Ba/Eu] ratios, as indicated in Figure \ref{fig:BaEu_dist}.  Figure \ref{fig:Patterns_s} shows the stars with $[\rm{Ba/Eu}]>-0.6$.  The top left panel of Figure \ref{fig:Patterns_s} shows the four stars with a suspected $s$-process contribution: these stars all have $[\rm{Ba/Eu}]>-0.4$, along with relatively high abundances of other elements, like La and Ce, that would reinforce the high Ba abundance.  The top right panel of Figure \ref{fig:Patterns_s} shows the stars with an unknown pattern: here the Ba abundance is high, but there are not enough elements to determine what the underlying pattern looks like.  The bottom two panels of Figure \ref{fig:Patterns_s} contain stars with $[\rm{Ba/Eu}]>-0.5$ (bottom left) and $-0.6<[\rm{Ba/Eu}]>-0.5$ (bottom right): these stars have high Ba abundances, yet the other abundances support an $r$-process pattern similar to Star 2792.  These bottom two panels indicate that the 4554 \AA \hspace{0.02in} line may be unreliable for determining Ba abundances in some stars---as such, the [Ba/Eu] ratio may not reflect the purity of the $r$-process signature.

Figure \ref{fig:Patterns_r} shows the stars with $[\rm{Ba/Eu}]<-0.6$.  The top left panel shows the lowest $[\rm{Eu/H}]$ stars; panels to the right and down show increasing $[\rm{Eu/H}]$ abundances.  These stars look to have a fairly robust pattern relative to 2792.  Though there are some small fluctuations in the patterns, they are generally within the uncertainties.  The exception is with Sr and Zr, which are occasionally slightly higher than 2792 \new{(see Section \ref{subsec:Spreads})}.

\subsection{Quantifying abundance spreads within M15}\label{subsec:Spreads}
In order to quantify the significance of the star-to-star spreads in neutron-capture elements, the spread ratio (SR) from \citet{Cohen2004} is utilized.  The SR is a ratio of the standard deviation of the sample, $\sigma$, to the typical uncertainty in a single abundance.  A larger SR indicates that the width of the distribution is significant compared to the uncertainty in an individual abundance; values of SR$>1$ indicate a significant spread.

Table \ref{table:SR_all} shows the mean abundances and spreads for all the M15 targets.  The SR$_{\rm{tot}}$ value shows the spreads amongst the actual abundances.  As expected, all the neutron-capture elements show a significant spread, while the spread in Fe is not significant.  The SR$_{\rm{Eu}}$ value shows the spreads once the abundances have all been shifted to the same Eu abundances (the patterns plotted in Figures \ref{fig:Patterns_s} and \ref{fig:Patterns_r}).  The SR$_{\rm{Eu}}$ values indicate that when shifted to a common Eu abundance, there are still significant spreads in Sr, Zr, Ba, Ce, and Nd, \new{likely due to the stars with a possible $s$-process signature.}

\begin{deluxetable}{@{}lccccc}
\newcolumntype{d}[1]{D{,}{\;\pm\;}{#1}}
\tabletypesize{\normalsize}
\tablecolumns{8}
\tablewidth{0pt}
\tablecaption{Average abundances and spread ratios of all 62 M15 stars\label{table:SR_all}}
\tablehead{
   & $\langle$[X/H]$\rangle$ & $N$ & $\sigma$ & SR$_{\rm{tot}}$\tablenotemark{a} & SR$_{\rm{Eu}}$\tablenotemark{b} }
\decimals
\startdata
Fe & -2.48 & 62 & 0.05 & 0.5  & --- \\
Sr & -2.58 & 10 & 0.13 & 1.31 & 1.20 \\
Zr & -2.07 & 60 & 0.13 & 1.31 & 1.30 \\
Ba & -2.41 & 62 & 0.30 & 2.96 & 2.13 \\
La & -1.86 & 34 & 0.15 & 1.46 & 0.87 \\
Ce & -2.12 & 45 & 0.16 & 1.63 & 1.03 \\
Nd & -1.88 & 42 & 0.19 & 1.86 & 1.26 \\
Sm & -1.76 & 35 & 0.16 & 1.56 & 0.83 \\
Eu & -1.74 & 62 & 0.20 & 2.02 & 0 \\
Dy & -1.52 & 32 & 0.16 & 1.57 & 0.59 \\
\enddata
\tablenotetext{a}{SR$_{\rm{tot}}$ is the spread ratio using the actual abundances.}
\tablenotetext{b}{SR$_{\rm{Eu}}$ is the spread ratio using the abundances when the Eu abundances have been shifted to the same value.}
\end{deluxetable}

Table \ref{table:SR_r} then shows the mean abundances and spreads only for the 43 stars with a confirmed $r$-process pattern, i.e., the ones shown in Figure \ref{fig:Patterns_r}.  Naturally, this selection has decreased the spread in Ba. The SR$_{\rm{Eu}}$ values show a consistent pattern amongst the lanthanides, though there are still significant spreads in Sr and Zr (however, note that there are fewer stars with Sr measurements).  This supports that the variations in the neutron-capture elements amongst these stars was due to an event that created these elements via the $r$-process pattern, creating a common abundance pattern in these 43 stars.

\begin{deluxetable}{@{}lccccc}
\newcolumntype{d}[1]{D{,}{\;\pm\;}{#1}}
\tabletypesize{\normalsize}
\tablecolumns{8}
\tablewidth{0pt}
\tablecaption{Average abundances and spread ratios of the 43 M15 stars with an $r$-process signature\label{table:SR_r}}
\tablehead{
   & $\langle$[X/H]$\rangle$ & $N$ & $\sigma$ & SR$_{\rm{tot}}$\tablenotemark{a} & SR$_{\rm{Eu}}$\tablenotemark{b} }
\decimals
\startdata
Sr & -2.58 &  7 & 0.15 & 1.48 & 1.21 \\
Zr & -2.09 & 41 & 0.13 & 1.26 & 1.19 \\
Ba & -2.53 & 43 & 0.22 & 2.22 & 0.77 \\
La & -1.87 & 25 & 0.15 & 1.50 & 0.70 \\
Ce & -2.14 & 32 & 0.17 & 1.65 & 0.69\\
Nd & -1.90 & 31 & 0.17 & 1.67 & 0.43 \\
Sm & -1.78 & 26 & 0.16 & 1.62 & 0.80 \\
Eu & -1.74 & 43 & 0.21 & 2.10 & 0 \\
Dy & -1.54 & 25 & 0.16 & 1.63 & 0.43 \\
\enddata
\tablenotetext{a}{SR$_{\rm{tot}}$ is the spread ratio using the actual abundances.}
\tablenotetext{b}{SR$_{\rm{Eu}}$ is the spread ratio using the abundances when the Eu abundances have been shifted to the same value.}
\end{deluxetable}

\begin{figure}[h!]
\begin{center}
\centering
\hspace*{-0.25in}
\includegraphics[scale=0.6,trim=0 0.in 0.in 0.0in,clip=True]{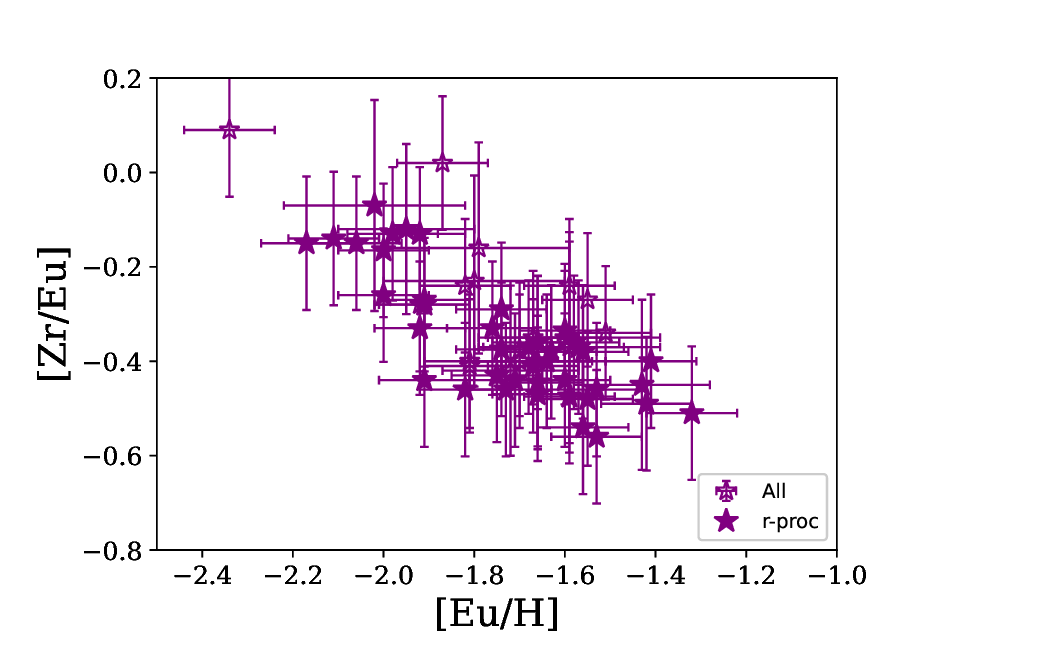}
\caption{The [Zr/Eu] ratios as a function of the [Eu/H] for all the stars in this paper. 
 The stars with a confirmed $r$-process pattern are shown as solid stars, while those with an uncertain pattern are shown as open stars.\label{fig:ZrEu}}
\end{center}
\end{figure}

\new{The Sr and Zr abundances show spreads independent of Eu and the other neutron-capture elements, as indicated in Tables \ref{table:SR_all} and \ref{table:SR_r}.  Along with Y, this difference was noticed previously by \citet{Otsuki2006} and \citet{Sobeck2011}, who both found an anti-correlation between [Zr/Eu] and [Eu/H].  The results from this paper also show this trend, as demonstrated in Figure \ref{fig:ZrEu}.  \citet{Otsuki2006} and \citet{Sobeck2011} both concluded that this correlation indicates that the lighter neutron-capture elements are produced in a different process from the heavier neutron-capture elements.  Figure \ref{fig:ZrEu} seems to support those conclusions: as an $r$-process event created more Eu, the [Zr/Eu] was lowered.  In their models, \citet{Tarumi2021} were further able to produce a similar result with lanthanide-rich ejecta from a NSM.  \citet{Kirby2023} similarly found a smaller spread of Sr, Y, and Zr compared to the heavier nuclides for M92.  }

\subsection{Classifications and distributions of stars}\label{subsec:RadialDist}
The majority of the target stars in M15 show a traditional $r$-process pattern.  However, most also meet the formal criterion for $r$-stars, $[\rm{Ba/Eu}] < 0$.  One AGB star, 54055, has an $s$-process signature and $[\rm{Ba/Eu}] > 0$---given its evolutionary status, this $s$-process signature is likely a result of dredge up or mass transfer from a companion. Overall, this analysis has identified 17 $r$-I stars, 44 $r$-II stars, 1 $s$-process star, and 1 star that falls just short of the $r$-I definition.  The latter star, 9753, may have an abundance pattern indicative of $s$-process contamination (see Figure \ref{fig:Patterns_s}).  Of the stars in the top panels of Figure \ref{fig:Patterns_s}, some would be classified as $r$-I stars, while others would be classified $r$-II stars, purely based on their Ba and Eu abundances.  However, as noted in Section \ref{subsec:Patterns}, the [Ba/Eu] may be an imperfect way of characterizing the $r$-process signature when using the 4554 \AA \hspace{0.02in} \ion{Ba}{2} line.

\begin{figure}[h!]
\begin{center}
\centering
\hspace*{-0.15in}
\includegraphics[scale=0.5,trim=0 0.in 0.in 0.0in,clip=True]{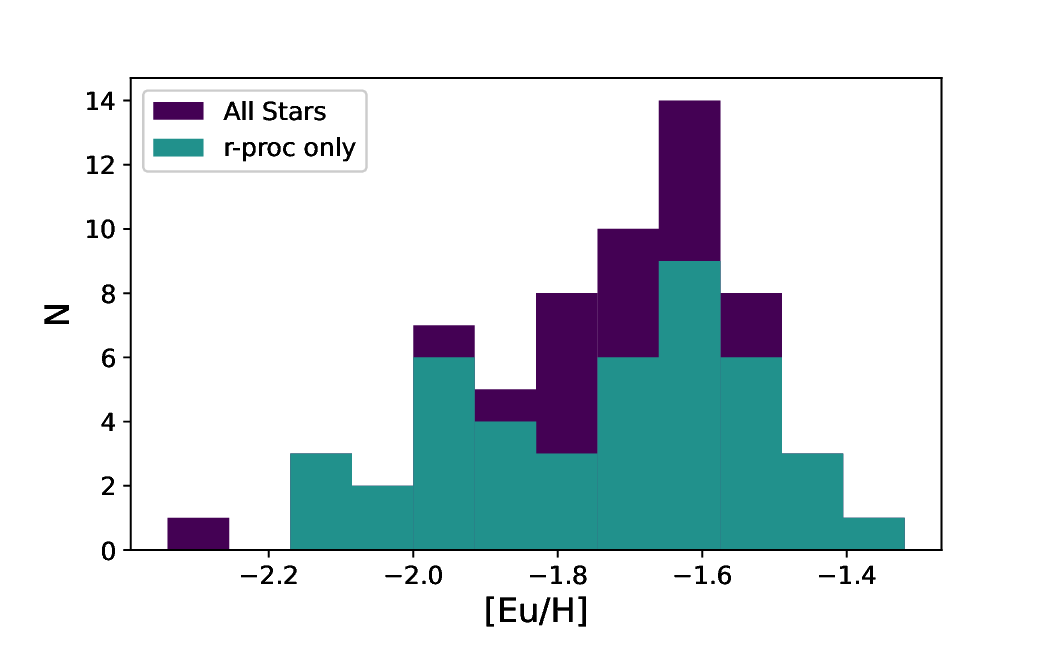}
\caption{The distribution of [Eu/H] abundances in M15 stars.\label{fig:EuH_Dist}}
\end{center}
\end{figure}

As found in previous papers \citep{Sneden1997,Sobeck2011,Worley2013}, M15 therefore has a dominant population of highly Eu-enhanced $r$-II stars.  Figure \ref{fig:EuH_Dist} shows the distribution of [Eu/H] abundances amongst the 62 targets in this paper \new{and for the 43 with a confirmed $r$-process pattern}.  \citet{Worley2013} found evidence for a bimodal distribution in [Ba/H].  While the distribution for all stars in Figure \ref{fig:EuH_Dist} is not bimodal, neither is it unimodal.  \new{The confirmed $r$-process stars do seem to show a bimodal distribution; a similar bimodality is also seen in [Eu/Fe].  The presence of a bimodality in [Eu/H] suggests two separate populations of stars: one that is moderately enhanced in Eu, typical of Milky Way field stars and other GCs, and another that is highly Eu-enhanced, similar to the halo $r$-II stars.  In their cosmological simulations, \citet{Tarumi2021} were able to reproduce the bimodality from \citet{Worley2013}, with requiring multiple epochs of star formation in M15; their resulting ``best-fit'' distribution looks very similar to the distribution in Figure \ref{fig:EuH_Dist}.  In M92, \citet{Kirby2023} found two discrete populations in Eu when the stars were separated by Na abundance, where the Eu spreads were confined to only the low-Na stars.  These relationships are explored further in M15 in Section \ref{subsec:Na}.}

\begin{figure*}[h!]
\begin{center}
\centering
\hspace*{-0.15in}
\subfigure{\includegraphics[scale=0.5,trim=0 0.3in 0.3in 0.75in,clip=True]{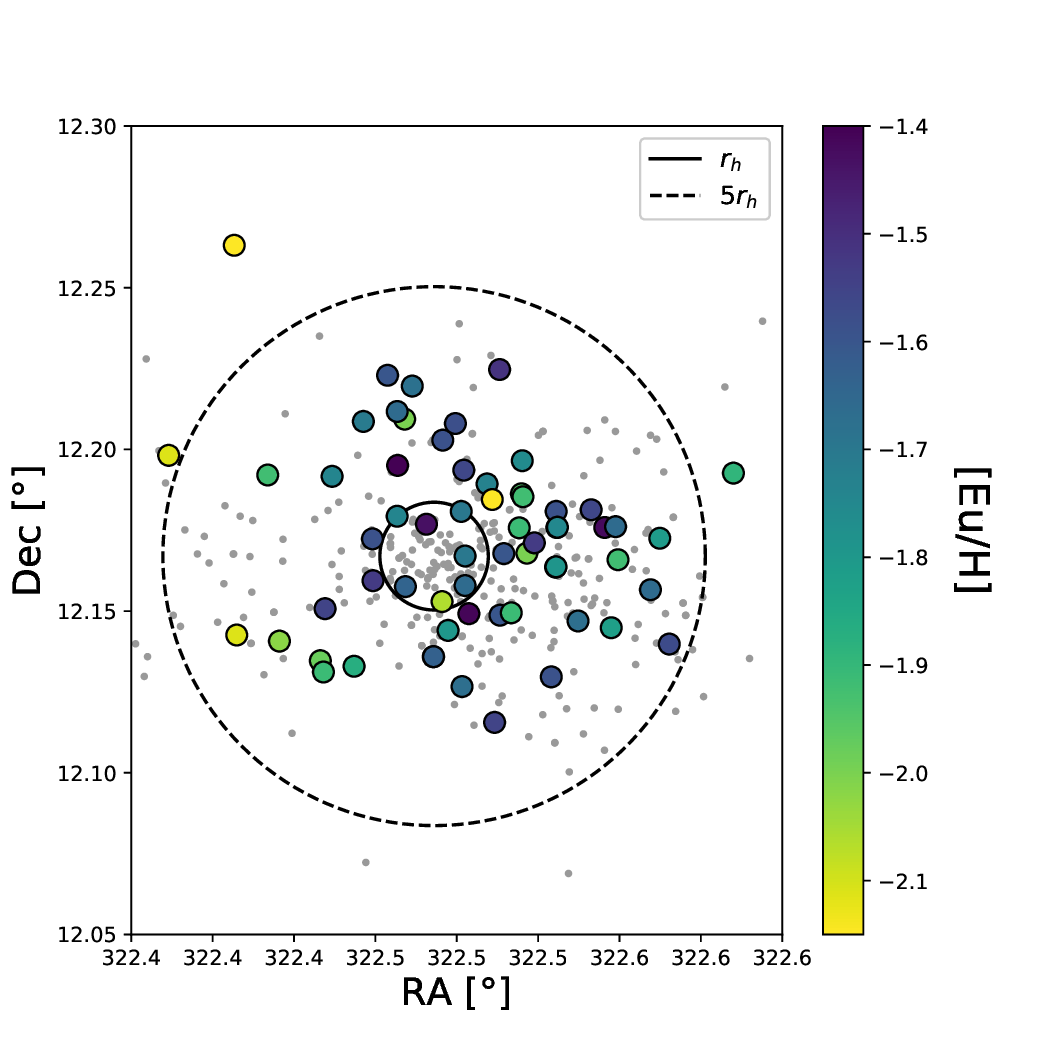}\label{subfig:Radial_Eu}}
\subfigure{\includegraphics[scale=0.5,trim=0 0.3in 0.3in 0.75in,clip=True]{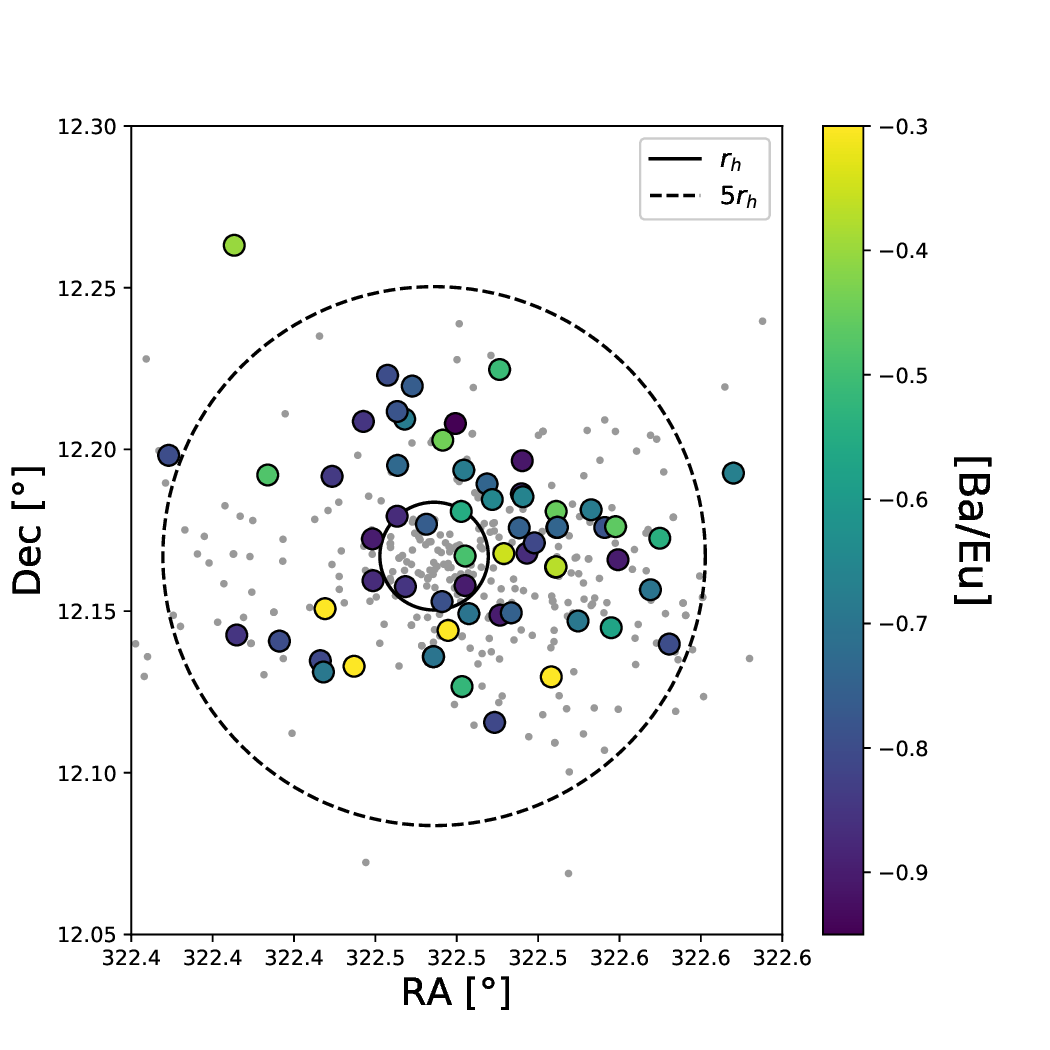}\label{subfig:Radial_BaEu}}
\caption{Positions of stars in M15.  The grey points show the stars from \citet{Kirby2016} and \citet{Carretta2009}, while the larger circles show the 62 star from this analysis.  For the \citet{Kirby2016} stars, note that the spatial distribution is heavily biased by slit selection constraints.  The points are color-coded by [Eu/Fe] (left) and [Ba/Eu] (right).  The solid circle shows the half-light radius \citep{Harris}, while the dashed circle shows five times the half-light radius.\label{fig:Radial}}
\end{center}
\end{figure*}

Figure \ref{fig:Radial} then investigates where in the cluster these stars lie.  The RA and Dec locations of the targets are plotted, along with the additional stars from \citet{Kirby2016} and \citet{Carretta2009}.  The stars are color-coded based on their [Eu/H] abundances (left) and [Ba/Eu] ratios (right). Based on these 2D projections, it is difficult to tell whether there are differences in the locations of stars with increased Eu abundances.  A larger sample of stars may be able to determine if there are significant differences.

\subsection{Relationships Between Neutron-Capture and Light Elements}\label{subsec:Na}
\begin{figure*}[h!]
\begin{center}
\centering
\hspace*{-0.15in}
\subfigure{\includegraphics[scale=0.64,trim=0.25in 0.in 1.25in 0.3in,clip=True]{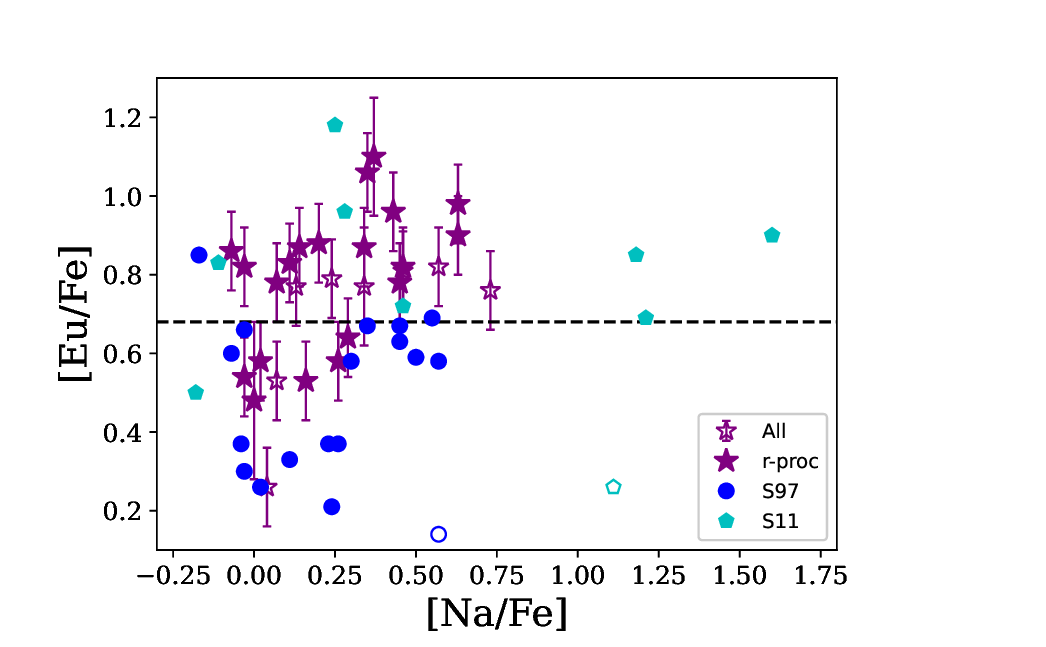}\label{subfig:Eu_Na}}
\subfigure{\includegraphics[scale=0.64,trim=0.15in 0.in 1.25in 0.3in,clip=True]{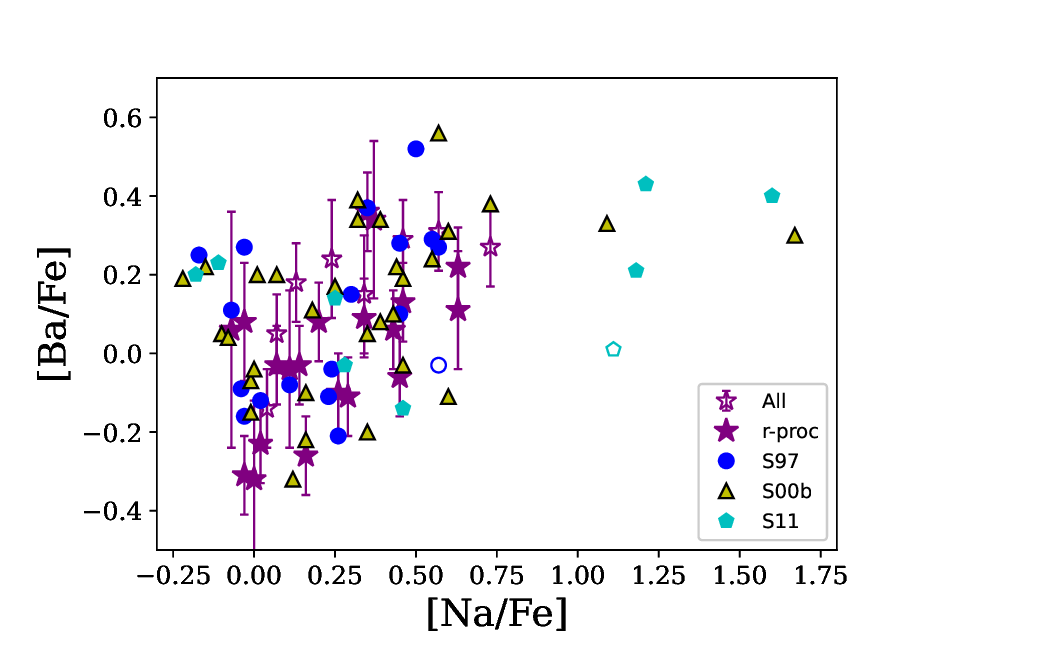}\label{subfig:Ba_Na}}
\subfigure{\includegraphics[scale=0.64,trim=0.15in 0.in 1.25in 0.3in,clip=True]{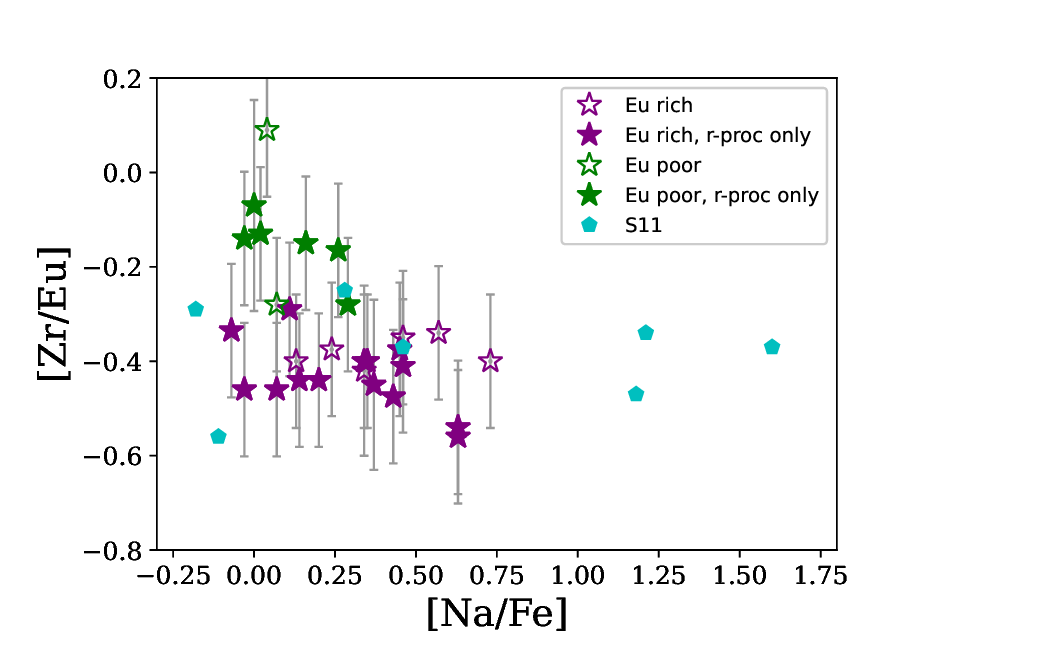}\label{subfig:ZrEu_Na}}
\caption{\new{Trends in neutron-capture abundances as a function of Na abundances.  The top panels show [Eu/Fe] and [Ba/Fe] as a function of [Na/Fe], while the bottom panel shows [Zr/Eu]  as a function of [Na/Fe].  The stars show the stars from this paper; filled are the stars with $r$-process patterns, while the open stars are stars with other patterns.  Na abundances for these stars are from \citet{Sneden1997,Sneden2000b,Carretta2009_GIRAFFE,Carretta2009,Sobeck2011}.  The top panels show points from the literature; the open point shows the outlier K583 from \citet{Sneden1997} and \citet{Sobeck2011}, as discussed in the text. 
 The dashed line in the [Eu/Fe] panel shows the division between low-Eu and high-Eu stars (see Figure \ref{fig:EuH_Dist}).  In the bottom panel, the Eu-rich and Eu-poor stars are identified in purple and green, respectively.}\label{fig:Na}}
\end{center}
\end{figure*}

\new{Previous attempts to explain the $r$-process element spreads in M15 and other GCs have used the relationship between the neutron-capture elements and Na as a way to probe the timing of the $r$-process event.  For instance, for M92, \citet{Kirby2023} recently found that the $r$-process element spreads within that GC were confined to the low-Na stars, while the more extreme, highly Na-enhanced stars were not found to host significant $r$-process spreads.  Several scenarios for creating Na and other light element spreads require two star formation events, where the low-Na stars would be the ``first generation'' of GC stars (see \citealt{BastianLardo2018} for a review of these scenarios, including pros and cons).  If the low-Na stars do represent a ``first generation'' of GC stars, the \citet{Kirby2023} result suggests that the NSM or other $r$-process event happened as the cluster's first stars were forming; the $r$-process-enhanced ejecta was then sufficiently mixed throughout the cluster before the formation of the second generation of stars.  The relationship between the neutron-capture elements and Na therefore places constraints on the timing and nature of the $r$-process event.  For instance, a NSM from the first generation of GC stars would only pollute the second generation (e.g., \citealt{Zevin2019}), the opposite of what was observed by \citet{Kirby2023} for M92.

For M15, a previous analysis by \citet{Roederer2011} found no relationship between La or Eu and Na, indicating that the $r$-process spreads were unrelated to the light element variations---however, the \citet{Roederer2011} analysis only utilized the 9 stars from \citet{Sobeck2011} that had both Na and Eu abudances determined in the same analysis. 
Unfortunately, there are no Na lines available in the M2FS spectral range in this paper.  In order to test the findings from \citet{Roederer2011}, [Na/Fe] ratios from the literature are utilized.  Of the 62 stars in this paper, 28 have previously-determined Na abundances from \citet{Sneden1997,Sneden2000b}, \citet{Carretta2009_GIRAFFE,Carretta2009}, and \citet{Sobeck2011}.  When stars overlapped between papers, the high-resolution analysis was preferred.  Note that for several stars, the Na abundances can vary significantly between papers; selecting alternate values does not have a significant effect on the findings below.

Figure \ref{fig:Na} shows the [Eu/Fe], [Ba/Fe], and [Zr/Eu] ratios as a function of the literature [Na/Fe] abundances.  The stars with an $r$-process pattern are indicated separately from those with an uncertain pattern.  Also shown are the literature values from \citet{Sneden1997,Sneden2000b} and \citet{Sobeck2011}, the papers which derived their own Na, Ba, and (except for \citealt{Sneden2000b}) Eu.  Figure \ref{fig:Na} hints at a trend in these abundance ratios with Na---Table \ref{table:Na} quantifies the strength of this correlation, using the Pearson correlation coefficient, as in \citet{Roederer2011}.  The table shows the number of stars considered for the correlation, $N$; the correlation coefficient, $r$, which quantifies how correlated ($r>0$) or anti-correlated ($r<0$) the two ratios are; and the probability that a random selection of $N$ stars would yield a correlation $\ge |r|$, $P_C(r,N)$.  Large values of $P_C(r,N)$ would indicate a high probability of a correlation being coincidental.}

\begin{deluxetable*}{@{}lcccc}
\newcolumntype{d}[1]{D{,}{\;\pm\;}{#1}}
\tabletypesize{\normalsize}
\tablecolumns{8}
\tablewidth{0pt}
\tablecaption{\new{Tests of correlations between neutron-capture and sodium abundance ratios.}\label{table:Na}}
\tablehead{
Sample & & [Ba/Fe]$+$[Na/Fe] & [Eu/Fe]$+$[Na/Fe] & [Zr/Eu]$+$[Na/Fe] }
\decimals
\startdata
This paper: all stars & & (0.62, 28, 0.00048) & (0.49, 28, 0.0087) & (-0.53, 28, 0.0037) \\
This paper: $r$-process only & & (0.56, 20, 0.0097) & (0.52, 20, 0.018) & (-0.59, 20, 0.0064) \\
& & & & \\
\citet{Sneden1997}\tablenotemark{a} & & (0.46, 17, 0.066) & (0.21, 17, 0.43) & --- \\
\citet{Sneden2000b} & & (0.40, 31, 0.026) & --- & --- \\
\citet{Sobeck2011}\tablenotemark{a} & & (0.52, 8, 0.19) & (0.11, 8, 0.79) & (-0.0080, 7, 0.99) \\
& & & & \\
This paper: Eu-rich stars & & (0.47, 20, 0.03) & (0.15, 20, 0.53) & (-0.28, 20, 0.22) \\
This paper: Eu-poor stars & & (0.44, 8, 0.28) & (0.46, 8, 0.25) & (-0.48, 8, 0.22) \\
& & & & \\
\enddata
Note: For each sample and abundance ratio combination, the quantities ($r$, $N$, $P_C(r,N)$ are given.  $N$ is the number of stars used to test the correlation; $r$ is the correlation coefficient, where a positive value indicates a correlation and a negative value indicates an anticorrelation; and $P_C(r,N)$ is the probability that a random sample of $N$ stars would give a correlation coefficient $\ge |r|$.
\tablenotetext{a}{K583 has been removed from this sample because of its uncertain abundances.}
\end{deluxetable*}

\new{Table \ref{table:Na} shows that for this paper and for the three literature samples, there is a modest, but significant correlation between [Ba/Fe] and [Na/Fe].  Note, however, that the strong 4554 \AA \hspace{0.015in} \ion{Ba}{2} line may be problematic, especially as the Ba abundances get high, as discussed in Section \ref{subsec:BaEu}.  For [Eu/Fe] and [Zr/Eu], a significant correlation and anticorrelation, respectively, are seen in this paper, but not in \citet{Sneden1997} or \citet{Sobeck2011}. Note that one star, K583, is a significant outlier in Figure \ref{fig:Na}, and has vastly different [Na/Fe] abundances between the two papers; for this reason, K583 is removed in the correlation tests in Table \ref{table:Na}.

Recall that a bimodality in [Eu/H] (and [Eu/Fe]) was found in Section \ref{subsec:RadialDist}.  When the stars are divided in Eu-rich and Eu-poor subsamples (see Figure \ref{fig:Na}), the correlations in Table \ref{table:Na} are affected.  The correlation in Ba is seen in both subsamples, but for [Eu/Fe] and [Zr/Eu] the correlations are only present in the Eu-poor samples.  This suggests that the Eu-poor stars are the ones driving the correlations, though the Eu-poor stars themselves are confined to lower [Na/Fe] ratios.  This difference between the Eu-rich and Eu-poor stars further explains the differences between this paper and the literature samples.  The \citet{Sneden1997} sample does not include many Na-rich stars and is primarily composed of Eu-poor stars: there is therefore not likely to be a strong correlation amongst these stars.  On the other hand, the \citet{Sobeck2011} sample is mainly composed of Eu-rich stars; the lack of correlations in their sample is consistent with the Eu-rich stars in this paper.

For M92, \citet{Kirby2023} did not find correlations between Na and neutron-capture elements, but rather a change in the spread of La and Eu between the Na-enhanced and low-Na stars.  The stars in Figure \ref{fig:Na} seem to hint at this trend: the Eu-poor stars have lower [Na/Fe] ratios, while the Eu-rich stars span a wide range of [Na/Fe] values, with little correlation in the [Eu/Fe] or [Zr/Eu] ratios in the Eu-rich population.  Unlike M92, M15's high-Na stars look to be Eu-enhanced.  Uncertainties in atmospheric parameters could create a correlation in abundance ratios, though no trends are seen in this paper.  Additional high-resolution spectroscopic follow-up will be necessary to quantify the spreads in the low- and high-Na subpopulations in M15, particularly with a sample that extends to higher Na values.

Ultimately, the results from this paper suggest that the $r$-process spreads are connected to the Na abundances, unlike what was previously found in the literature.}

\section{Conclusion}\label{sec:Conclusion}
This paper has presented neutron-capture abundances for 62 stars in the globular cluster, M15.  The spectra were obtained with the M2FS spectrograph and cover a relatively small wavelength range from about 4430-4630 \AA.  This wavelength coverage provides spectral lines from \ion{Fe}{1}, \ion{Fe}{2}, \ion{Sr}{1}, \ion{Zr}{2}, \ion{Ba}{2}, \ion{La}{2}, \ion{Ce}{2}, \ion{Nd}{2}, \ion{Sm}{2}, \ion{Eu}{2}, and \ion{Dy}{2}.  The main findings of this paper are summarized below.
\begin{itemize}
\item The [Fe/H] ratios derived in this analysis are found to be within the range found in the literature.
\item Abundances of the target stars were calculated relative to the high S/N star, 2792.  This star is found to be an $r$-II star with a pattern of neutron-capture elements that is similar to the $r$-process pattern in the Sun.
\item The derived [Eu/H] abundances were found to be in general agreement with values from the literature.  The majority of the target stars (44) are found to be highly Eu-enhanced $r$-II stars, while another 17 are moderately Eu-enhanced $r$-I stars.  One star is consistent with an $s$-process signature, based on its [Ba/Eu] ratio.  The [Eu/H] distribution is found to be inconsistent with a Gaussian distribution; \new{the stars with a confirmed $r$-process pattern amongst the lanthanides show a bimodality in [Eu/H].}
\item The Ba abundances derived from the 4554 \AA \hspace{0.02in} line are occasionally in disagreement with typical values from the literature.  Moreover, ten stars from this analysis are found to have high Ba abundances, relative to Eu and the other lanthanides.  This suggests that this strong Ba line may not be suitable for determining Ba abundances or [Ba/Eu] ratios for these M15 stars.  However, four stars are found to have high Ba, La, Ce, and Nd ratios that support contamination from the $s$-process.  Another five stars only have Zr, Ba, and Eu ratios, such that their neutron-capture abundances could not be determined.
\item The 62 target stars are found to show significant star-to-star spreads in Sr, Zr, Ba, La, Ce, Nd, Sm, Eu, and Dy, but {\it not} in Fe.  When the high Ba stars are removed, the spreads are still significant amongst the 43 remaining stars.  When the abundances are shifted so that the Eu abundances are identical, there are no significant spreads in Ba, La, Ce, Nd, Sm, or Dy.  This suggests that the 43 stars in M15 were enhanced by the same process, and that the nucleosynthetic source of this Eu pollution was the $r$-process.
\item \new{When Na abundances from the literature are included, the stars show correlations between the neutron-capture elements ([Ba/Fe], [Eu/Fe], and [Zr/Eu]) and [Na/Fe], contrary to what was previously found by \citet{Roederer2011}.  This analysis finds that the Eu-rich cluster stars cover a wide spread in [Na/Fe], while lower Eu stars are confined to low Na.  These results appear to be consistent with the recent M92 results from \citet{Kirby2023}, suggesting that the $r$-process spreads are limited to the low-Na population of stars.  A larger, high-resolution survey of M15 stars is needed to investigate further.}
\end{itemize}

\new{Ultimately, the results from this paper are consistent with models that require an $r$-process nucleosynthetic event to occur early on, as the first cluster stars are forming (e.g., \citealt{Tarumi2021}).  Such an early event may pose a challenge for NSMs (see, e.g., \citealt{Kobayashi2023}).  Additional follow-up observations of M15 are needed to further characterize the nature of the $r$-process and light element spreads.}

\acknowledgements 
The authors thank the anonymous reviewer for suggestions that have improved this manuscript.  JCG and CMS thank Evan Kirby for discussions and suggestions that have improved this manuscript.
CMS acknowledges support from the National Science Foundation grant AST 2206379.
D.E. acknowledges support through the NIH-RISE scholarship (R25-GM059298).

\software{%
IRAF (Tody 1986, Tody 1993), 
DAOSPEC \citep{Stetson2008}, 
MOOG (v2017; \citealt{Sneden1973,Sobeck2011}), 
linemake (\url{https://github.com/vmplacco/linemake})
}

\footnotesize{}

\end{document}